\documentclass[
 reprint,
 superscriptaddress,
 amsmath,amssymb,
 aps,
 prb,twocolumn,
 showkeys
]{revtex4-2}

\usepackage{graphicx}
\usepackage{dcolumn}
\usepackage{bm}
\usepackage{graphicx,epstopdf}
\usepackage{tikz}
\usepackage{textcomp,gensymb}
\usepackage[export]{adjustbox}
\usepackage{stackengine}
\usepackage{natbib}
\usepackage{xcolor}
\usepackage{hyperref}
\hypersetup{
	colorlinks=true,
	linkcolor=blue,
	urlcolor=blue,
	citecolor=blue
}
\usepackage{placeins}
\usepackage{glossaries}

\newacronym[plural=magnetoresistances]{mr}{MR}{magnetoresistance}
\newacronym{nlmr}{NLMR}{negative longitudinal magnetoresistance}
\newacronym[plural=transition-metal dichalcogenides]{tmd}{TDM}{transition-metal dichalcogenide}
\newacronym[plural=Weyl points]{wp}{WP}{Weyl point}
\newacronym[plural=Dirac points]{dp}{DP}{Dirac point}
\newacronym{vdw}{vdW}{van-der-Waals}
\newacronym{cme}{CME}{chiral magnetic effect}
\newacronym{trs}{TRS}{time-reversal symmetry}
\newacronym{is}{IS}{inversion symmetry}
\newacronym{xps}{XPS}{X-ray photoemission spectroscopy}
\newacronym{pdms}{PDMS}{Polydimethylsiloxane}
\newacronym{ipa}{IPA}{isopropyl alcohol}
\newacronym{rt}{R-T}{resistance-over-temperature-curve}
\newacronym[plural=field-coolings]{fc}{FC}{field-cooled}
\newacronym[plural=zero-field-coolings]{zfc}{ZFC}{zero-field-cooled}
\newacronym{fw}{FW}{field-warming}
\newacronym{dft}{DFT}{density functional theory}
\newacronym{2d}{2D}{two-dimensional}
\newacronym{dos}{DOS}{density of states}
\newacronym{wal}{WAL}{weak anti-localization}
\newacronym{lowess}{LOWESS}{locally weighted scatterplot smoothing}
\newacronym{sh}{SH}{sample holder}

\newcommand{\pt}{$\text{PtSe}_2\,$}

\newcommand{\fig}{Fig.~}
\newcommand{\figs}{Figs.~}
\newcommand{\eq}{Eq.~}

\newcommand{\tab}{Tab.~}
\newcommand{\refText}{Ref.~}

\newcommand{\figureS}{\setcounter{figure}{0}\renewcommand{\thefigure}{S\arabic{figure}}}
\newcommand{\tableS}{\setcounter{table}{0}\renewcommand{\thetable}{S\arabic{table}}}
\newcommand{\equationS}{\setcounter{equation}{0}\renewcommand{\theequation}{S\arabic{equation}}}

\begin{document}
\preprint{APS/123-QED}
\title{Negative Longitudinal Magnetoresistance in the\\ Dirac Semimetal \pt - Kondo Effect and Surface Spin Dynamics}
\author{Julian Max Salchegger}
\email{julian.salchegger@jku.at}
\affiliation{Institut f\"ur Halbleiter-und-Festk\"orperphysik, Johannes Kepler University, Altenbergerstr. 69, A-4040 Linz, Austria}
\author{Rajdeep Adhikari}
\email{rajdeep.adhikari@jku.at}
\affiliation{Institut f\"ur Halbleiter-und-Festk\"orperphysik, Johannes Kepler University, Altenbergerstr. 69, A-4040 Linz, Austria}
\affiliation{Linz institute of Technology, Johannes Kepler University, Altenbergerstr. 69, A-4040 Linz, Austria}
\author{Bogdan Faina}
\affiliation{Institut f\"ur Halbleiter-und-Festk\"orperphysik, Johannes Kepler University, Altenbergerstr. 69, A-4040 Linz, Austria}
\author{Jelena Pešić}
\affiliation{Institute of Physics Belgrade, University of Belgrade, Pregrevica 118, 11080 Belgrade, Serbia}
\author{Alberta Bonanni}
\email{alberta.bonanni@jku.at}
\affiliation{Institut f\"ur Halbleiter-und-Festk\"orperphysik, Johannes Kepler University, Altenbergerstr. 69, A-4040 Linz, Austria}
\date{\today}

\begin{abstract}
The emergence of negative longitudinal magnetoresistance in the topologically non-trivial transition-metal dichalcogenide \pt is studied. Low $T$/high $\mu_0 H$ transport is performed for arbitrary field directions, and an analytical framework is established. The source of the negative longitudinal magnetoresistance is identified as the Kondo effect stemming from Pt-vacancies contributing an uncompensated spin, exclusively at the sample surface. The concentration of vacancies and the sample thickness are identified as tuning parameters. The findings are substantiated by density functional theory, which corroborates the proposed Pt-vacancy model.
\end{abstract}

\keywords{\pt, transition-metal dichalcogenide, Kondo effect, vacancy, transport, negative longitudinal magnetoresistance}
\maketitle

\section{\label{intro_root}Introduction}
Beyond graphene \cite{GeimNovoselov}, \glspl{tmd} (\textit{e.g.} \pt, $\text{NiTe}_2\,$, $\text{PdTe}_2\,$, $\text{WTe}_2\,$) support a plethora of exploitable features, such as the \gls{cme} \cite{WangWTe2CME,LiPtSe2NLMR}, thickness-dependent semiconductor-to-semimetal transition \cite{Villaos2019}, type-II Dirac cones breaking the Lorentz invariance \cite{DasWTe2Overview} and topologically non-trivial surface states \cite{GhoshNiTe2Edge}.

The presence of topological features is a consequence of \glspl{tmd} possessing a Weyl- or Dirac-semimetallic nature, which opens avenues towards prospective computing practices \cite{KononovWTe2Edge,ZhangWSe2photodet} and a window to high energy physics phenomena through the lens of quasiparticles at low temperatures \cite{BurkovWeylPrediction,ZhouIsingSCMajorana,ChenAxion}. Specifically, the materials host at least one pair of \glspl{wp} in their Brillouin zone. The \glspl{wp} own a chiral charge of $\pm 1$, with the total chiral charge over the whole Brillouin zone vanishing. The Weyl- and Dirac-semimetals can be distinguished according to symmetry: A Dirac-semimetal hosts at least one \gls{dp}, and preserves both the \gls{trs} and \gls{is}, while for a Weyl-semimetal at least one of these symmetries is broken and a pair of \glspl{wp} emerges \cite{Armitage}. A \gls{dp} can then be described as consisting of a pair of \glspl{wp} with opposite chirality at the same momentum, thus the \gls{dp} is not chiral. A linear dispersion relation $E \propto |\boldsymbol{k}|$, is found to connect the charge carrier energy $E$ with the momentum $\boldsymbol{k}$ around the \gls{dp} or \gls{wp}, wherein charge carriers have zero effective mass. This linear band dispersion around a \gls{dp} generates a Dirac cone. If the dispersion is anisotropic with respect to $\boldsymbol{k}$, \textit{i.e.} it is linear only along certain directions, the Dirac cone is type-II-tilted. In this case, the resulting surface states (boundary to trivial topology) are also anisotropic.

This occurs in \pt, hosting \glspl{dp} located at the $\overline{\mathit{\Gamma} A}$-axis, around which the respective Dirac cones are type-II-tilted with respect to the selfsame $\overline{\mathit{\Gamma} A}$-axis. Strictly speaking, the \glspl{dp} are only established in the bulk limit. Monolayers and bilayers of \pt are semiconducting \cite{Villaos2019,KandemirThinPtSe2}, while a transition to a semimetallic behavior takes place for a thickness of three monolayers. Bulk \pt is AA-stacked and is $C_3$-symmetric with respect to the out-of-plane axis $\overline{\mathit{\Gamma} A}$. This symmetry protects the accidental band crossing forming the type-II \glspl{dp}. The projection of the bulk \glspl{dp} onto the boundary to trivial topology is predicted to result in helical surface states along the edges of the crystal for the Fermi level being sufficiently close to the \gls{dp} energy level \cite{Armitage, MehdiArcProtec}. These topological characteristics, combined with high electronic mobility, significant spin-orbit-coupling, and stability under ambient conditions, make semimetallic \pt a top-contender for employment in electronic \cite{YimPtSe2Electric}, spintronic \cite{AbsorPtSe2SOC}, as well as in photonic \cite{WuPhotonic} devices. Recent developments also point to \pt as a candidate for the experimental observation of the orbital Hall effect \cite{SahuOrbitalHall, CrPS4PtHetero}.

Here, low $T$/high $\mu_0 H$ transport in \pt is carried out in order to gain insight into the electronic properties of the system. In particular, the origin of the \gls{nlmr} detected at temperatures $T \leq 2.5\,\text{K}$ \cite{LiPtSe2NLMR, JieweiPtSe2chiralAnomaly}, is explored. The presence of a Kondo effect is identified, which is heralded by a global non-zero-temperature resistance-minimum and a \gls{nlmr}. The origin and the observed thickness-dependence of the Kondo effect are elucidated \textit{via ab initio} calculations.

\section{\label{exp_root}Experimental}
\begin{figure}
    \includegraphics[width=0.6\linewidth]{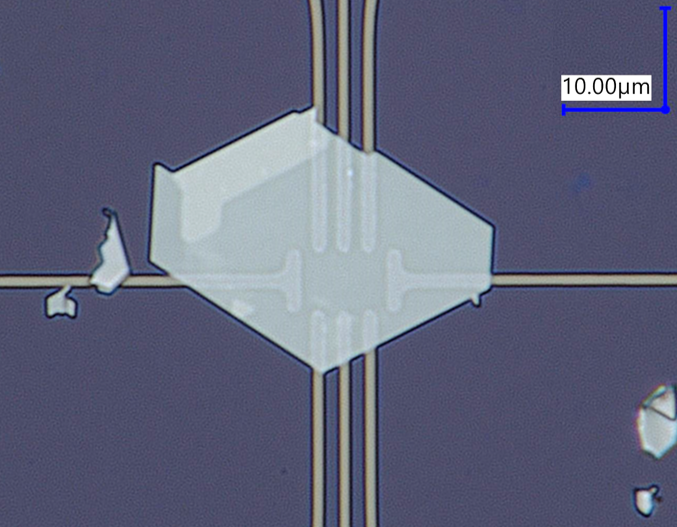}
    \caption{Optical image of sample B.}
    \label{i_OM}
\end{figure}

The \pt flakes considered in this work are obtained by mechanical exfoliation from a bulk crystal grown by HQ Graphene \cite{HQGraphene}, instead of by direct selenation of Pt. This ensures high crystallinity and long-range-order, as confirmed by optical microscopy, shown in \fig \ref{i_OM}, along with atomic-force microscopy and Raman spectroscopy, depicted in \figs \ref{i_AFM} and \ref{i_Raman} respectively in the Supplemental Material. The optical microscopy shows that the angles between flake edges are comprised of $n\cdot60\degree, n\in\mathbb{N}$. Atomic force microscopy additionally resolves the flake facets as flat. Raman spectroscopy confirms the expected structural phase. Three representative \pt samples A, B and C with thicknesses in the range of $20\,\text{nm}$ are listed in \tab \ref{t_samples}. The selected $20\,\text{nm}$ regime provides a bulk bandstructure while allowing a spread of surface-to-bulk ratios from $18\,\text{nm}$ to $26\,\text{nm}$. The flakes are placed onto Pt-contacts, which are patterned \textit{via} electron-beam lithography. Sample A has a 6-terminal Hall-bar design with a $(3\times5)\,\mu\text{m}^2$ measuring area. Sample B has a 8-terminal Hall-bar design with a $(5\times7)\,\mu\text{m}^2$ measuring area. An optical image of sample B is shown in \fig \ref{i_OM}. The superseding 8-terminal design is introduced because of symmetry considerations for terminal pairs normal to source-drain and increased fault-tolerance \textit{via} redundancy. For sample C a design identical to the one of sample B is chosen.

\begin{table}
    \begin{tabular}{c|c|c}
        sample & thickness $t$ (nm) & Hall-bar design\\ \hline
        A      & 18             & 6-terminal\\ 
        B      & 20             & 8-terminal\\ 
        C      & 26             & 8-terminal\\ 
        \end{tabular}
    \caption{Considered \pt-samples together with thickness and Hall-bar geometry.}
    \label{t_samples}
\end{table}

\begin{figure}
\includegraphics[width=0.95\linewidth]{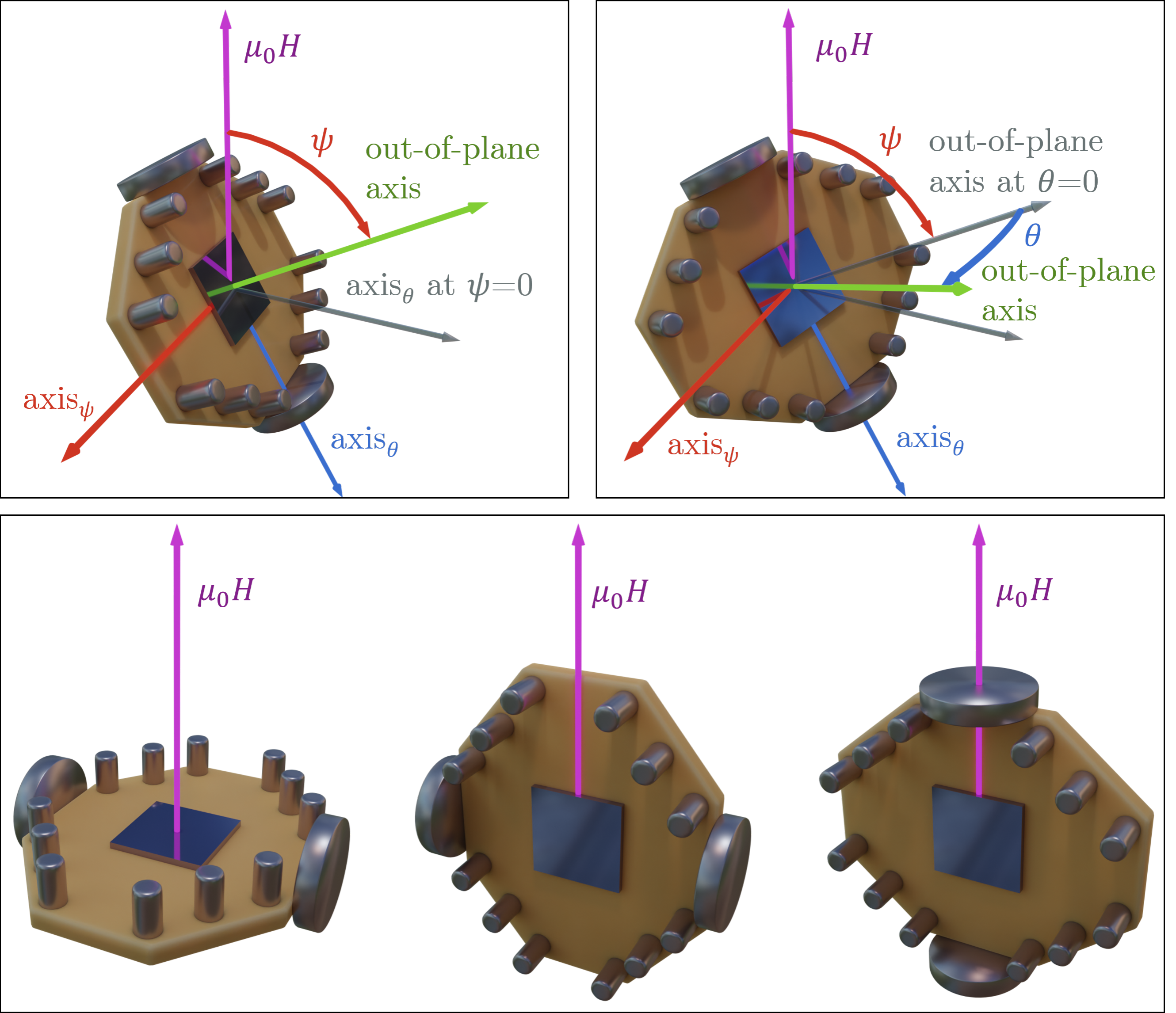}
\stackinset{c}{-0.92\linewidth}{c}{0.78\linewidth}{a)}{\hspace{0pt}}
\stackinset{c}{-0.46\linewidth}{c}{0.78\linewidth}{b)}{\hspace{0pt}}
\stackinset{c}{-0.95\linewidth}{c}{0.37\linewidth}{c)}{\hspace{0pt}}
\caption{a) Model of the \gls{sh} showing the rotation-axes $\text{axis}_\theta$ and $\text{axis}_\psi$ and the respective angles $\theta$ and $\psi$, as well as the applied magnetic field and the \gls{sh} out-of-plane axis. a) A rotation around $\text{axis}_\psi$ puts the \gls{sh} into a $(0,\psi)$ orientation. b) An additional rotation around $\text{axis}_\theta$ results in a $(\theta, \psi)$ orientation. c) $(0,0)$ orientation (left), $(90,0)$ (middle) and $(90,90)$ (right).}
\label{i_SH}
\end{figure}

Finally, Au wires are bonded to the Pt-contact-pads using In as a solder agent. The low $T$/high $\mu_0 H$ transport measurements are then carried out in a Janis Super Variable Temperature 7TM-SVM cryostat equipped with a 7\,T superconducting magnet and an indigenously developed rotatory \gls{sh} with two angular degrees of freedom. A lock-in amplifier \textit{ac} technique is employed for measuring the magnetotransport properties of the exfoliated \pt flakes. The \gls{sh} allows rotation along the $\text{axis}_\psi$, a static axis normal to $\boldsymbol{H}$ (vector of applied magnetic field), and $\psi$ is the angle measured between the \gls{sh} out-of plane axis and $\boldsymbol{H}$ (\fig \ref{i_SH}\,a)). An additional rotation along $\text{axis}_\theta$ is available, with $\text{axis}_\theta$ being normal to the $\text{axis}_\psi$ and rotating with the \gls{sh}. The resulting angle $\theta$ is measured between the out-of-plane axis of the \gls{sh} and the plane spanned by $(\boldsymbol{H}\wedge\text{axis}_\theta)$, depicted in \fig \ref{i_SH}\,b). The orientation of the \gls{sh} is thus defined by $(\theta, \psi)$ with $\theta$ and $\psi$ given in degrees. In \fig \ref{i_SH}\,c), the orientations $(0,0)$, $(90,0)$, $(90,90)$ are depicted (left to right). The specification of $\theta$ and $\psi$ differs from the conventional designation of spherical coordinates by the fact that the axis$_\theta$ itself rotates along with $\psi$. For samples with \gls{is}, such as \pt, $\psi \rightarrow -\psi$ and $\theta \rightarrow -\theta$ are expected to be unity operations. For sample A, the rotation is monitored using the sample’s Hall voltage as well as that of a graphite Hall bar mounted at an angle of $90\,\degree$ with respect to the \gls{sh} out-of-plane axis. For samples B and C, the orientation is established \textit{via} increments driven by the stepper motors responsible for the $\psi$- and $\theta$-rotations.

\section{\label{res_root}Results and Discussion}
\subsection{\label{res_RT}Temperature-dependence of the resistance}
\begin{figure}[htb]
\includegraphics[width=\linewidth]{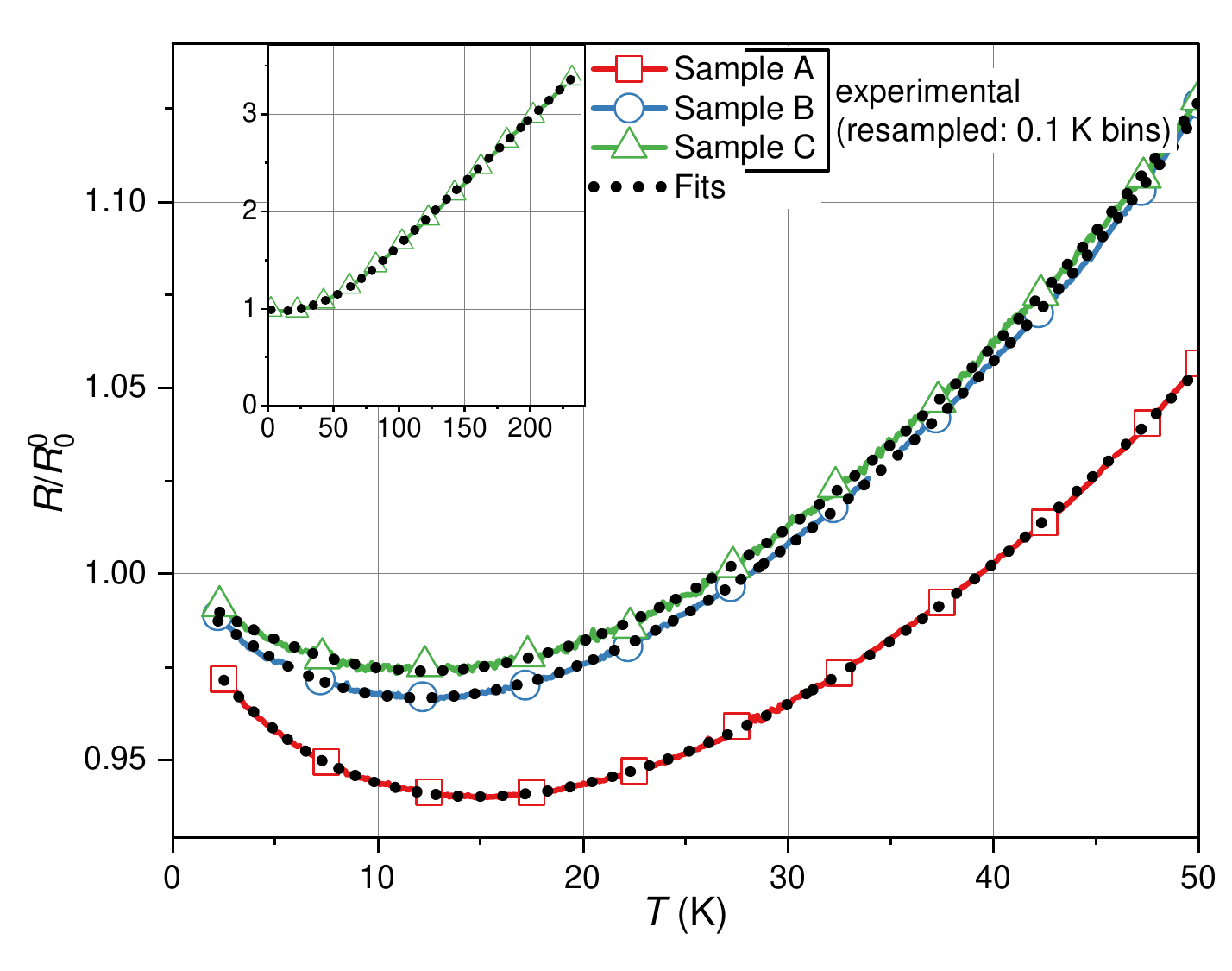}
\caption{Resistance ratio $\frac{R}{R_0^0}$ over temperature at zero magnetic field for the considered samples. The respective fits are shown as dotted lines. Inset: The resistance increases {$\propto T$ for $T>T_\text{lin}\approx 80\,\text{K}$}.}
\label{i_ZFC}
\end{figure}

To obtain the longitudinal resistance, voltage probing terminals are chosen on the same side of the drain-source centerline of the Hall-bar. \Gls{zfc} measurements are carried out by cooling the sample in the absence of a magnetic field. The \gls{zfc} behavior of the resistance as a function of temperature for the three samples is shown in \fig \ref{i_ZFC}: The resistance ratio to the theoretical zero-field zero-temperature resistance, $R/R_0^0$, initially decreases with temperature, as shown in the inset, and a minimum-resistance is found at $T\sim 12\,\text{K}$ for all samples. As the temperature is further reduced, the resistance increases again. The data is resampled into $0.1\,\text{K}$ bins (evaluated as average of all data points inside the bin). A larger temperature-range for sample C is provided in the inset to \fig \ref{i_ZFC}.

This characteristic is the consequence of two features: (1) the semimetallic nature of \pt gives rise to a Fermi-liquid behavior and a lattice vibration term at low temperatures, (2) a Kondo effect causes an increase in resistance for $T\rightarrow 2\,\text{K}$. The Kondo effect describes scattering processes in which the charge carrier spin is exchanged with the spin of a magnetic impurity (twice), altering the wave vector of the charge carrier. This leads to a logarithmic correction to the resistance. A minimum-resistance temperature is obtained for antiferromagnetic (\textit{i.e.} negative) coupling of the charge carrier spin to the impurity spin \cite{Kondo}. The resistance is not assumed to diverge for $T\rightarrow 0$ because the impurity spins are screened  by interacting charge carriers for $T\lesssim T_\text{K}$, with $T_\text{K}$ being the Kondo temperature. The Kondo effect acts also in the case of magnetic moments resulting from vacancies, and Pt-vacancies are indeed observed in \pt \cite{AvsarThinPtSe2, Ge}.

The semimetallic resistance $R_\text{SM}$ accounting for the Fermi liquid behavior (electron-electron coupling) and the electron-phonon coupling is given by:
\begin{equation}
    R_\text{SM}(T) = R_\text{res}+A_\text{F} T^2 + A_\text{ph} T^5,\quad T\leq T_\text{lin}.
    \label{eq_RSM_lowT}
\end{equation}
Where $T_\text{lin} \approx 80\,\text{K}$ is the transition temperature between the quadratic Fermi-liquid behavior plus the $\propto T^5$ lattice vibration term to the linear regime. The theoretical residual resistance at $T=0$ (without the contribution of the Kondo effect) is $R_{\text{res}}$, and $A_\text{F}$ and $A_\text{ph}$ are prefactors of the Fermi liquid- and the lattice vibration resistance, respectively. For $T_\text{K} \ll T_\text{lin}$, the linear regime is not relevant for the analysis of the Kondo effect. The expression for $R_\text{SM}(T)$ at arbitrary temperatures is given in \eq \ref{eq_RSM_full} in the Supplemental Material. 

Analytical descriptions of the Kondo effect are elusive and often valid only over a limited range of parameters \cite{CostiKondo}. In particular, temperature ranges around $T\approx T_\text{K}$ are challenging to treat. An expression for the Kondo resistance is adapted from \refText \cite{Hamann}, in that the prefactor $\frac{1}{2} R_\text{K0}$, with $R_\text{K0}$ the (maximum) Kondo resistance, is introduced to account for the nonlocal resistance differing from the local resistance by a constant factor. The original prefactors are contained in $R_\text{K0}$. The expression for the Kondo resistance becomes then:
\begin{equation}
    R_\text{K}(T) = R_\text{K0}\frac{1}{2}\left(1-\frac{\ln({T/T_\text{K}})}{\sqrt{\ln({T/T_\text{K}})^2+S(S+1)\pi^2}}\right).
    \label{eq_RK}
\end{equation}
Here, The Kondo temperature is given by $T_\text{K}$ and $S$ is the impurity spin. Strictly speaking, $S$ is not a good quantum number, since it models the spin of hybrid orbitals and should thus not be understood as an actual spin quantum number.

The zero-field resistance $R(H=0,T)$ is obtained as
\begin{equation}
    R(H=0,T) = R_\text{SM}(T)+R_\text{K}(T),
    \label{eq_ZFC}
\end{equation}
and is employed to fit the obtained \gls{zfc} data, which is shown in \fig \ref{i_ZFC} as dotted lines. The total resistance at zero field and zero temperature is given by $R_0^0 = R(H=0,T=0) = R_{\text{res}}+R_{\text{K0}}$ and $H$ is the magnitude of the magnetic field $\left|\boldsymbol{H}\right|$. The observation of a global resistance minimum $R_\text{min}$ is an indication of the onset of the Kondo effect. As the thickness $t$ of the samples increases ($t_A < t_B < t_C$), the Kondo resistance $R_\text{K0}$ decreases, which can be explained by considering the system as a bulk conductance channel, except for the surface, where Pt-vacancies form uncompensated spins which yield a Kondo resistance. The retrieved values of $S$ are given in \tab \ref{t_ZFC_param} in the Supplemental Material and are comparable to those reported in \refText \cite{Ge}.

\subsection{\label{res_MR}Magnetic field dependence of the resistance}
\begin{figure}[htb]
\includegraphics[width=\linewidth]{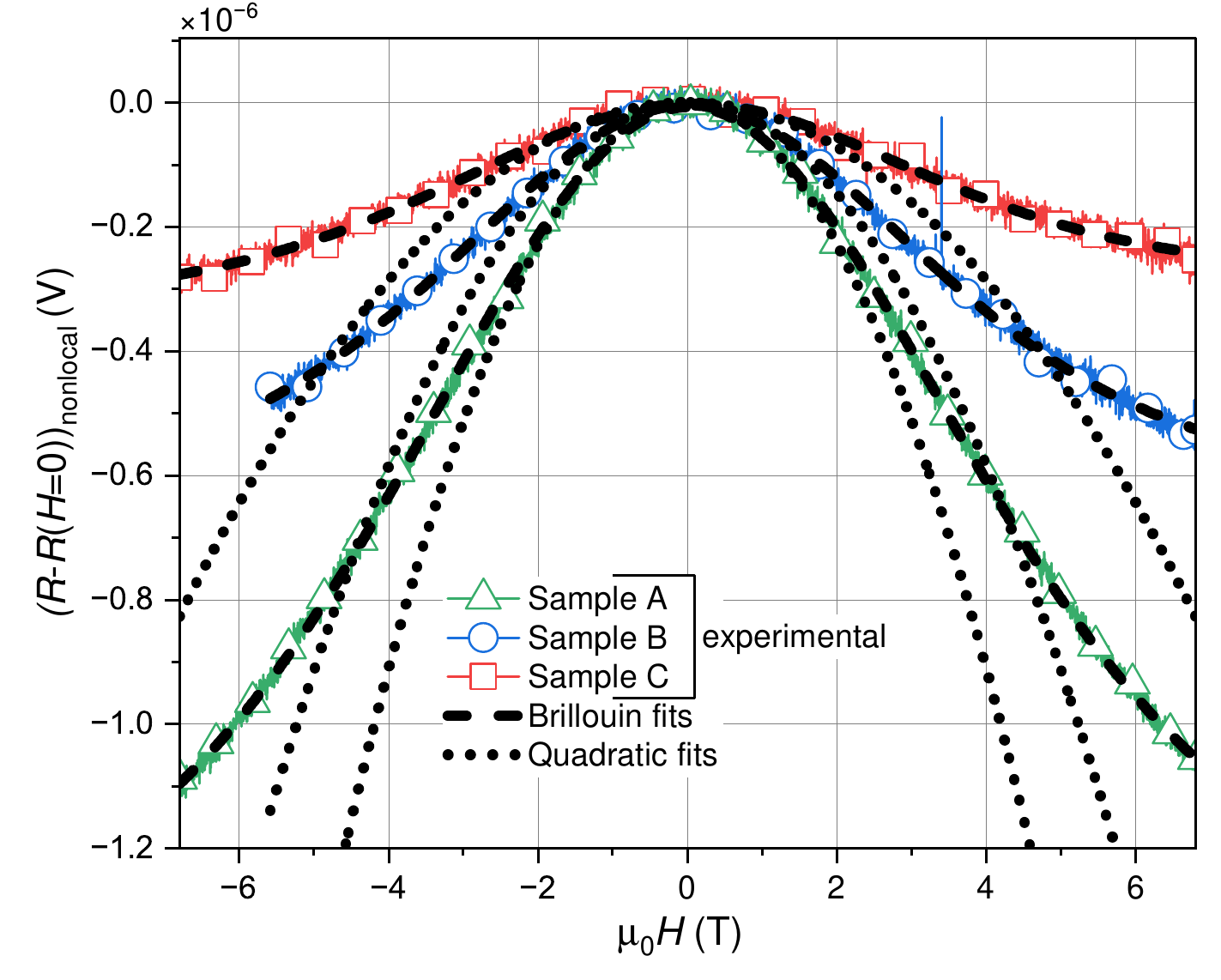}
\caption{Resistance $(R-R(H=0))$ over applied magnetic field for all samples in the $(90,0)$ orientation, isolating the Kondo effect from any orbital magnetoresistance. Dotted lines: quadratic fits. Dashed lines: fits according to \eq \ref{eq_felschfactor}.}
\label{i_MR_90_0}
\end{figure}

The (longitudinal) \gls{mr} is probed by choosing voltage probing terminals at the same side of the drain-source centerline, but keeping the temperature constant while sweeping the magnetic field. A \gls{nlmr} is observed for all samples when taking the \gls{mr} in the $(90,0)$ orientation, as depicted in \fig \ref{i_MR_90_0}, which shows the resistance $(R-R(H=0))$ decreasing for increasing $H$. The onset of saturation is observed for $\mu_0 H>4\,\text{T}$. The relative magnitude of the \gls{nlmr} is greatest for the thinnest sample A, and diminishes with sample thickness.

Since in this orientation there are no magnetic field components normal to the electric field $\boldsymbol{E}$, orbital magnetoresistance can be ruled out. The \gls{nlmr} is attributed to the Kondo effect: By applying a magnetic field, the spin-flip scattering, which is the cause of the Kondo resistance, is inhibited \textit{via} polarization of the impurity spins and of the charge carrier spins. The slope of $R(H,T)$ plateaus out for $H \rightarrow \infty$, due to saturation of the polarization. This effect is assumed to take place isotropically for any direction of $\boldsymbol{H}$.

Without detailed knowledge about the orbital nature of the uncompensated spins at the surface of the \pt flake and thereby of the coupling of the vacancy spins with the conduction electrons or their magnetization $\boldsymbol{M}(\boldsymbol{H})$, the magnetic field dependence of the Kondo resistance $R_\text{K}(H,T)$ cannot be described rigorously at arbitrary temperatures and fields. Analytical descriptions are challenged in the regime where $H$ induces a partial, but non-negligible polarization of the impurities.

A quadratic dependence $R(H) \propto -\alpha (\mu_0 H)^2, \alpha>0$ is obtained by approximating a low-spin-density-expansion around $H=0$ \cite{AndreiKondoSolution, Ge}, which can be fitted satisfactorily to the data for $\mu_0 H < 2\,\text{T}$. The fit is shown as the dotted lines in \fig \ref{i_MR_90_0}.

In order to take the magnetization saturation into account, the model has to be modified for magnetic fields higher than $\mu_0 H \approx 2\,\text{T}$. Thus, \eq \ref{eq_RK} is complemented by an impurity spin polarization factor, which accounts for the charge carrier spins aligning with the impurity spins. At $H \to \pm\infty$, the polarization saturates \cite{Felsch, BealMonodWeiner}:
\begin{equation}
    R_\text{K}(H,T) = R_\text{K}(T) \left( 1-\mathfrak{B}_J\left(\frac{g \mu_B \mu_0 H}{k_B (T+T_\text{K})}\right)^2\right),
\label{eq_felschfactor}
\end{equation}
with $\mathfrak{B}_J$ being the Brillouin function for the total angular momentum magnitude $J=\frac{1}{2}$, $g$ is the Landé $g$-factor, $\mu_B$ the Bohr magneton and $\mu_0$ the vacuum permeability. It can be seen from the Brillouin fits in \fig \ref{i_MR_90_0} (dashed lines), that this model reasonably fits the data, accounting for the required saturation behavior.

A \gls{cme} (or Adler-Bell-Jackiw anomaly) may also be considered as source of the \gls{nlmr}, as previously reported for \pt \cite{LiPtSe2NLMR, JieweiPtSe2chiralAnomaly}. In the context of solid-state physics, a \gls{cme} describes the generation of a chiral current, usually in Weyl- and Dirac-semimetals \cite{Armitage, BurkovWeylPrediction}. In the case of Dirac-systems, the \gls{dp} has net-zero chirality and the breaking of symmetries is required for the \gls{cme} to occur: if the \gls{trs} is broken \textit{via} the application of an external magnetic field $\boldsymbol{H}$, the chiral degeneracy of the (quadruply degenerate) \gls{dp} is lifted, and the \gls{dp} splits along $\boldsymbol{H}$ into a pair of (doubly degenerate) \glspl{wp}. By applying an electric field $\boldsymbol{E}$ parallel to $\boldsymbol{H}$, a chemical potential difference $\Delta\mu$ is established between the \glspl{wp}, effectively pumping chiral charge carriers by moving them from one \gls{wp} to the one of opposite chirality. The system still possesses net-zero chirality, and the chiral charge conservation leads to the chiral charge carriers having to relax at the opposite \gls{wp} (or \textit{via} scattering). Resulting from the chemical potential difference $\Delta\mu$ between the \glspl{wp}, a chiral current density emerges. Since the chiral charge carriers have an electric charge, an electric current density establishes concurrently to the chiral one, resulting in a \gls{nlmr}.
This effect requires components of $\boldsymbol{E} \parallel \boldsymbol{H}$ and the chiral current density $\boldsymbol{J_C}$ diminishes according to:
\begin{equation}
\left|\boldsymbol{J_C}\right| \propto \left(\cos\angle(\boldsymbol{E},\boldsymbol{H})\right)^2 .
\end{equation}

Because the obtained \gls{nlmr} in the examined samples is robust with respect to $\angle(\boldsymbol{E},\boldsymbol{H})$ and persists even at the $(\pm90,\pm90)$ orientations ($\boldsymbol{E} \bot \boldsymbol{H}$ and $\boldsymbol{H}$ in-plane), a \gls{cme} does not conform to the acquired data. The observed modulation of the Kondo effect by sample thickness can also not be understood in the frame of a \gls{cme}. A contribution to the already-negative \gls{mr} in the $\boldsymbol{E} \parallel \boldsymbol{H}$ configuration is not be excluded, but it cannot be the dominant effect causing the detected \gls{nlmr} and cannot explain the increase in resistance observed for $T\rightarrow 0$.

Further, it is unclear whether the type-II \gls{dp} in \pt could show a reasonable contribution to the conductivity, since the \gls{dp} is found $\approx1.2\,\text{eV}$ below the Fermi energy $E_\text{F}$ \cite{HuangPtSe2BANDDFTandARPES}. From the calculated band structures, type-I \glspl{dp} above $E_\text{F}$ could also be considered \cite{HuangPtSe2BANDDFTandARPES, KenanPtSe2BANDDFTandARPES}. It was shown, that the \gls{cme} conductance is $\propto (E_\text{F}-E_{\text{DP}})^{-2}$ \cite{SpivakCME}, with $E_{\text{DP}}$ the energy at the \gls{dp}. This lessens the need for \glspl{dp} close to $E_\text{F}$ in order to obtain a significant conductivity contribution. Another option is to consider defects which modify $E_\text{F}$, potentially bringing it closer to a \gls{dp} and again, increasing the \gls{cme} contribution to the conductivity. It is therefore realistic, that \pt can host a \gls{cme}. 

Since the \gls{mr} remains negative even in the $(\pm90,\pm90)$ orientations, current jetting \cite{YangCurrentJet} is an unlikely source of \gls{nlmr}, since it should scale with $\propto (\cos \angle(\boldsymbol{E},\boldsymbol{H}))^2$. Additionally, the well-defined Hall-bar geometry should not enable current jetting at arbitrary angles. Furthermore, current jetting explains neither the saturation of the \gls{nlmr} for $H\gtrsim 4\,\text{T}$, nor the increase in resistance observed for $T\rightarrow 0$.

In the $(0,0)$ orientation, the \gls{mr} is positive for all samples and only varies slightly over the measured temperatures for $T\leq 50\,\text{K}$. The resistance $(R-R(H=0))$ over applied magnetic field in $(0,0)$ orientation is given in \fig \ref{i_MR_0_0_A} for sample A and in \fig \ref{i_MR_0_0_BC} in the Supplemental Material for samples B and C.

\begin{figure}
    \includegraphics[width=0.9\linewidth]{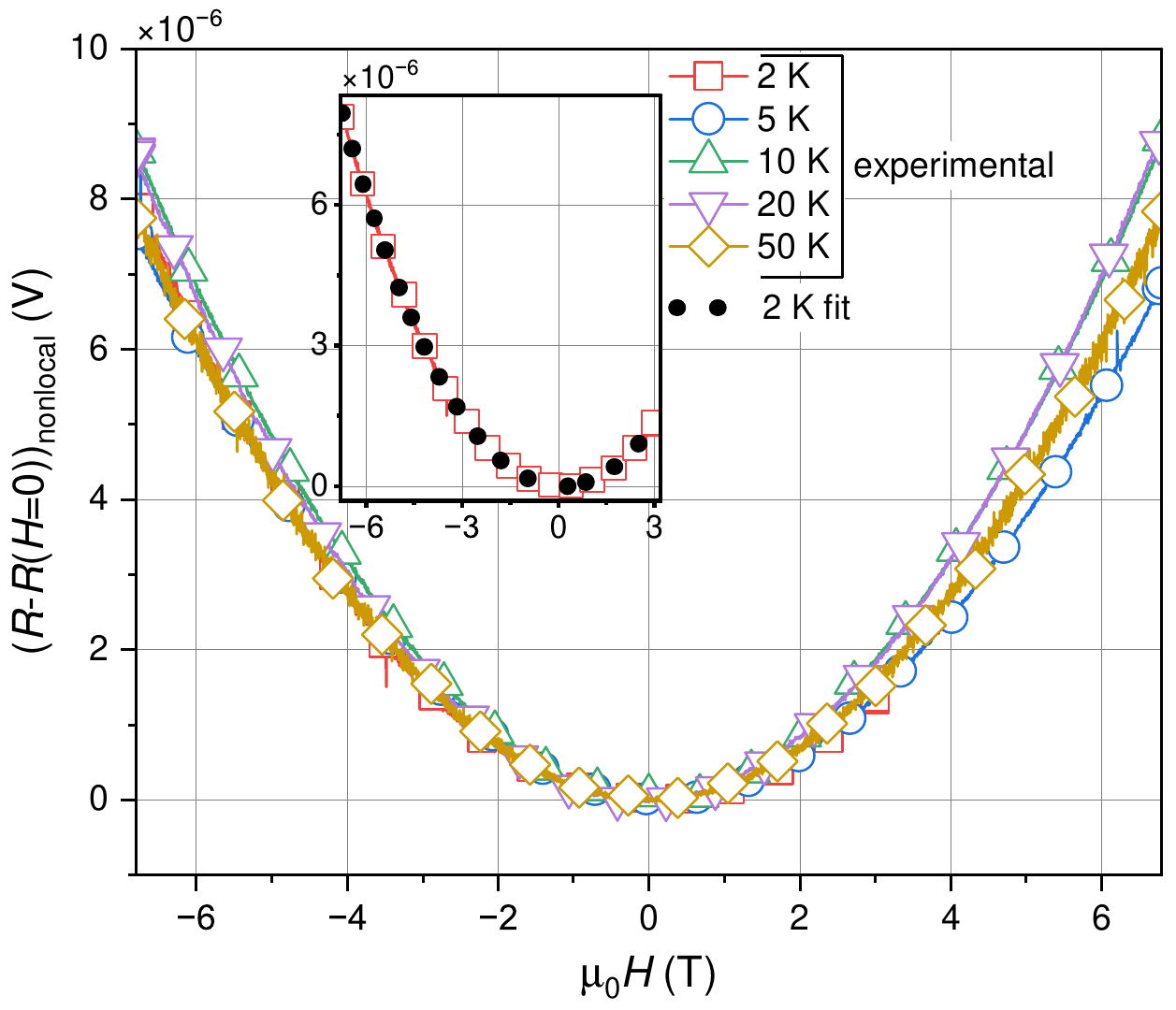}
    \caption{Resistance $(R-R(H=0))$ of sample A over applied magnetic field in $(0,0)$ orientation over a temperature range of $(2-50)\,\text{K}$. Inset: Fit (dotted line) at $2\,\text{K}$.}
    \label{i_MR_0_0_A}
\end{figure}

The observed evolution of the resistance is $\approx (R\propto H^2)$, pointing at a system dominated by orbital \gls{mr}. The weak temperature-dependence can be explained by the fact, that all data lie in the Fermi-liquid regime for $T < T_\text{lin} \approx 80\,\text{K}$. The magnetic field components $\boldsymbol{H} \parallel \boldsymbol{E}$ do not contribute any orbital \gls{mr}, given that the resulting Lorentz force is zero. Due to the finite sample thickness and to the crystalline anisotropy, components of $\boldsymbol{H}$ out-of-plane and $\boldsymbol{H}$ in-plane give rise to distinct orbital \glspl{mr} effects. It is worth noting, that because of the isotropic nature of the Kondo effect, the orbital \gls{mr} cannot be singled out in the temperature range around $T_\text{K}$.

\begin{figure}
    \includegraphics[width=0.9\linewidth]{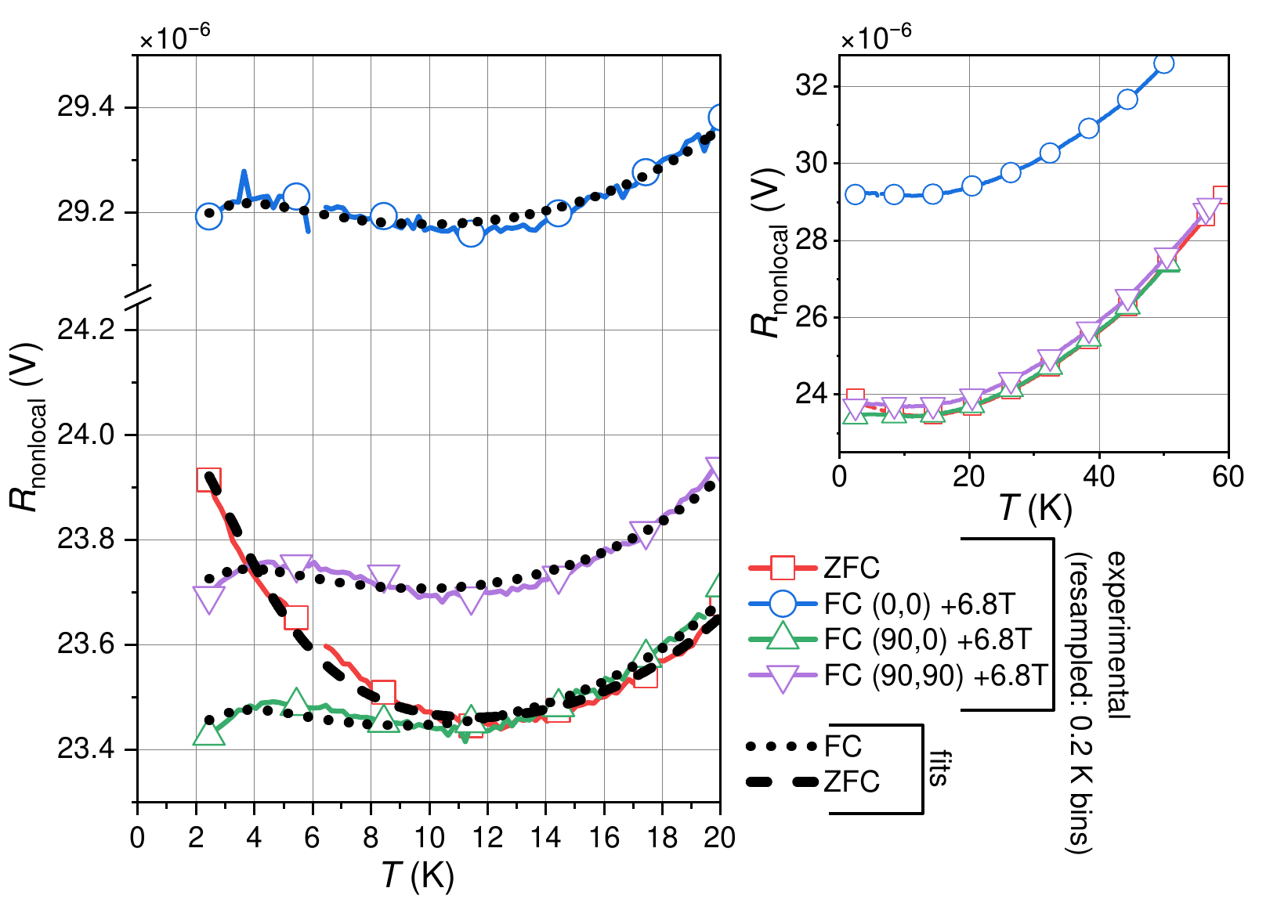}
    \stackinset{c}{-0.92\linewidth}{c}{0.61\linewidth}{a)}{\hspace{0pt}}\\
    \includegraphics[width=0.95\linewidth]{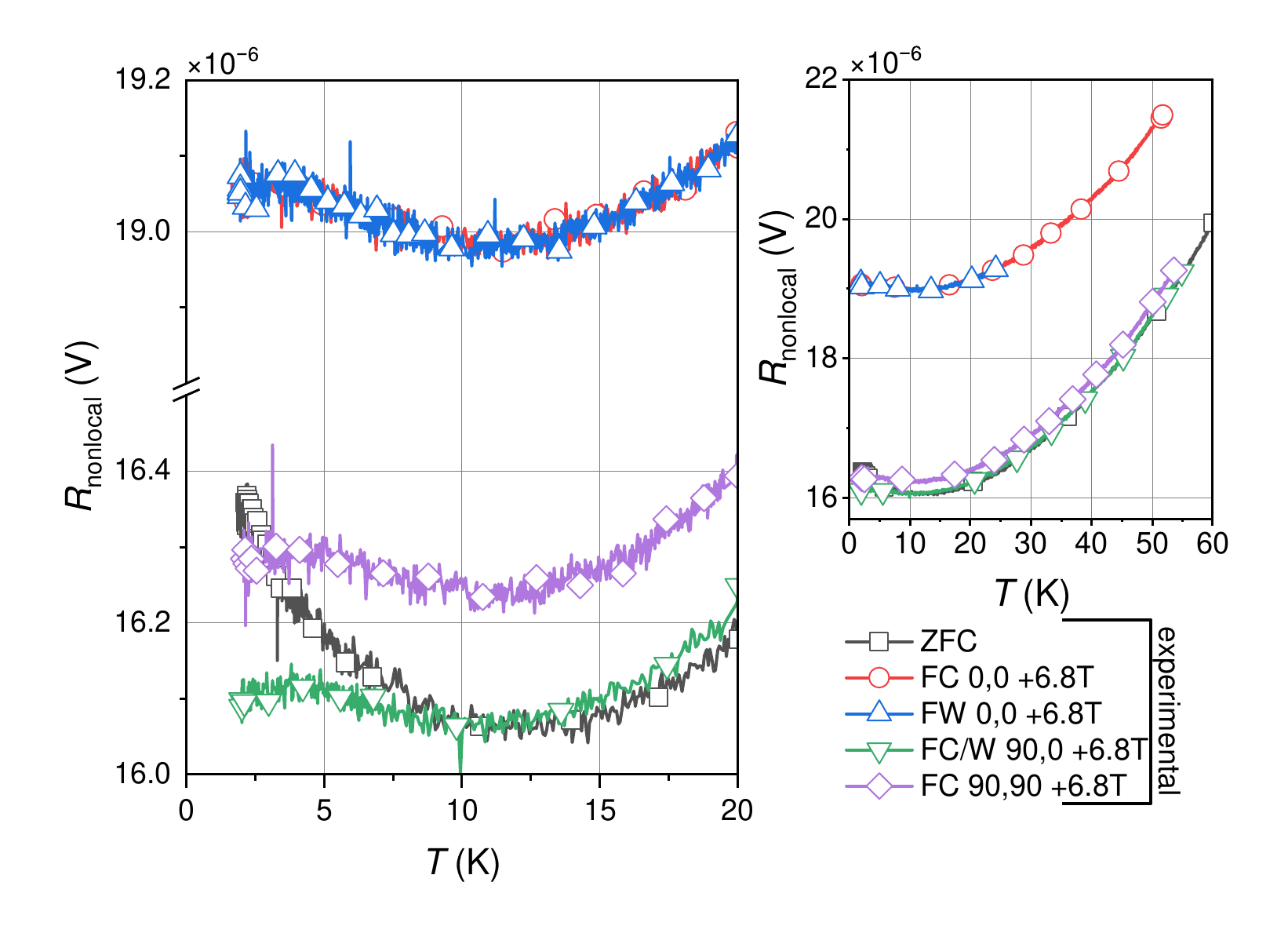}
    \stackinset{c}{-0.92\linewidth}{c}{0.61\linewidth}{b)}{\hspace{0pt}}
    \caption{Resistance over temperature curves (\glspl{zfc} and \glspl{fc}/\gls{fw}). a) Sample B. Left panel: Over a temperature range of $T<20\,\text{K}$, including an axis-break. Dotted lines: Fit of the \glspl{fc}. Dashed lines: Fit of the \gls{zfc}. Right panel: Over a temperature range of $T<60\,\text{K}$. b) Sample C. Left panel: Over a temperature range of $T\leq20\,\text{K}$, including an axis-break. The \gls{fw} in $(0,0)$ orientation is indistinguishable from the respective \gls{fc}. Right panel: Over a temperature range of $T<60\,\text{K}$.}
    \label{i_FCs}
\end{figure}

The orbital \glspl{mr} thus require separate fitting parameters for the orientations
\begin{equation}
    \mathfrak{o} \in\{\bot,\parallel\}=\{\text{out-of-plane},\,\text{in-plane}\}.
\end{equation}
This treatment is similar to the introduction of an anisotropic electron mass \cite{ZhaoguoPtSe2MRAni}. The description of the orbital \gls{mr} $R_{\text{orb},\mathfrak{o}}$ in orientation $\mathfrak{o}$, is given by: 
\begin{equation}
    \label{eq_MR0}
    R_{\text{orb},\mathfrak{o}}(H) = M_\mathfrak{o} \left|\mu_0 H\right| ^{p_\mathfrak{o}} + k_\mathfrak{o} \mu_0 H \quad \mathfrak{o}\in\{\bot,\parallel\}.
\end{equation}
Here, the parameter $M_\mathfrak{o}$ models the magnitude of the \gls{mr}, $p_\mathfrak{o}$ the power with which the \gls{mr} scales ($p_\mathfrak{o}=2$ for a Fermi-liquid) and $k_\mathfrak{o}$ is a linear correction term to account for temperature gradients and imperfect geometry of the contacts to the flake. The fit for the total resistance of sample A in the $(0,0)$ orientation at $T=2\,\text{K}$, shown in the inset of \fig \ref{i_MR_0_0_A}, follows:
\begin{equation}
    R(H) = \left(R_\text{SM}(T)+R_\text{K}(H,T)\right)\big|_{T=2\,\text{K}}+R_{\text{orb},\mathfrak{o}}(H) 
\end{equation}
with $R_\text{SM}(T)+R_\text{K}(H,T)$ determined by fitting the field cooling of sample A in the $(-90,0)$ orientation (which has no orbital effect components). The weak temperature-dependence of $R_{\text{orb},\mathfrak{o}}$ and the reduced magnitude of $M_\parallel$ \textit{versus} $M_\bot$ can also be gleaned from \fig \ref{i_FCs}\,a),b), where the \gls{rt} curves for samples B and C are given.

\subsection{Unified transport model}
\begin{figure}
\includegraphics[width=\linewidth]{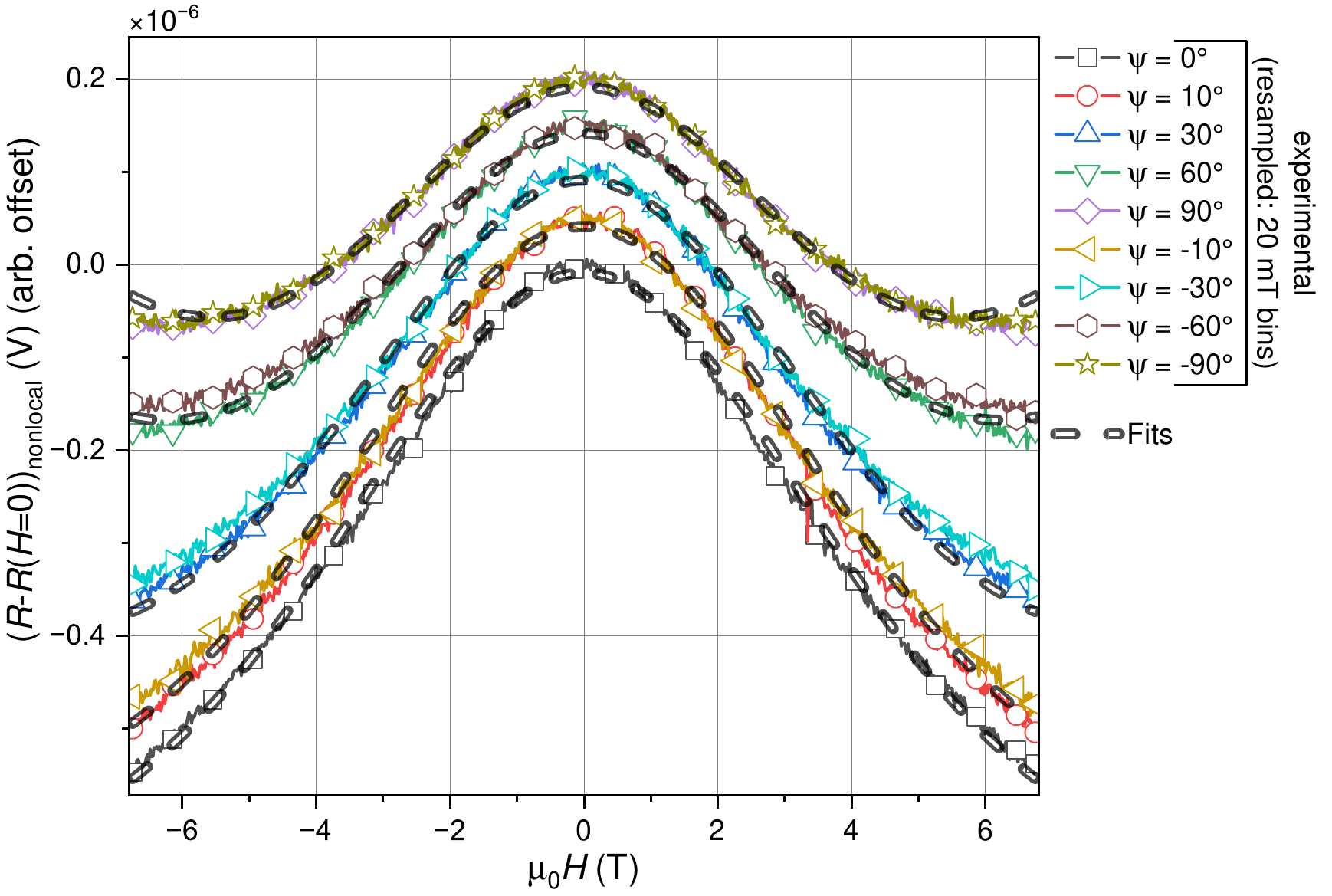}
\caption{Resistance over applied magnetic field curves of sample B at $2\,\text{K}$ for orientations $\left(90\,\degree,(-90\,\degree \leq \psi \leq 90\,\degree)\right)$, including fitting-functions. Arbitrary offset. Data is resampled every $20\,\text{mT}$ and linear components are disregarded to better compare it to the fit.}
\label{i_MR90_psi}
\end{figure}

\begin{figure}
\includegraphics[width=0.85\linewidth]{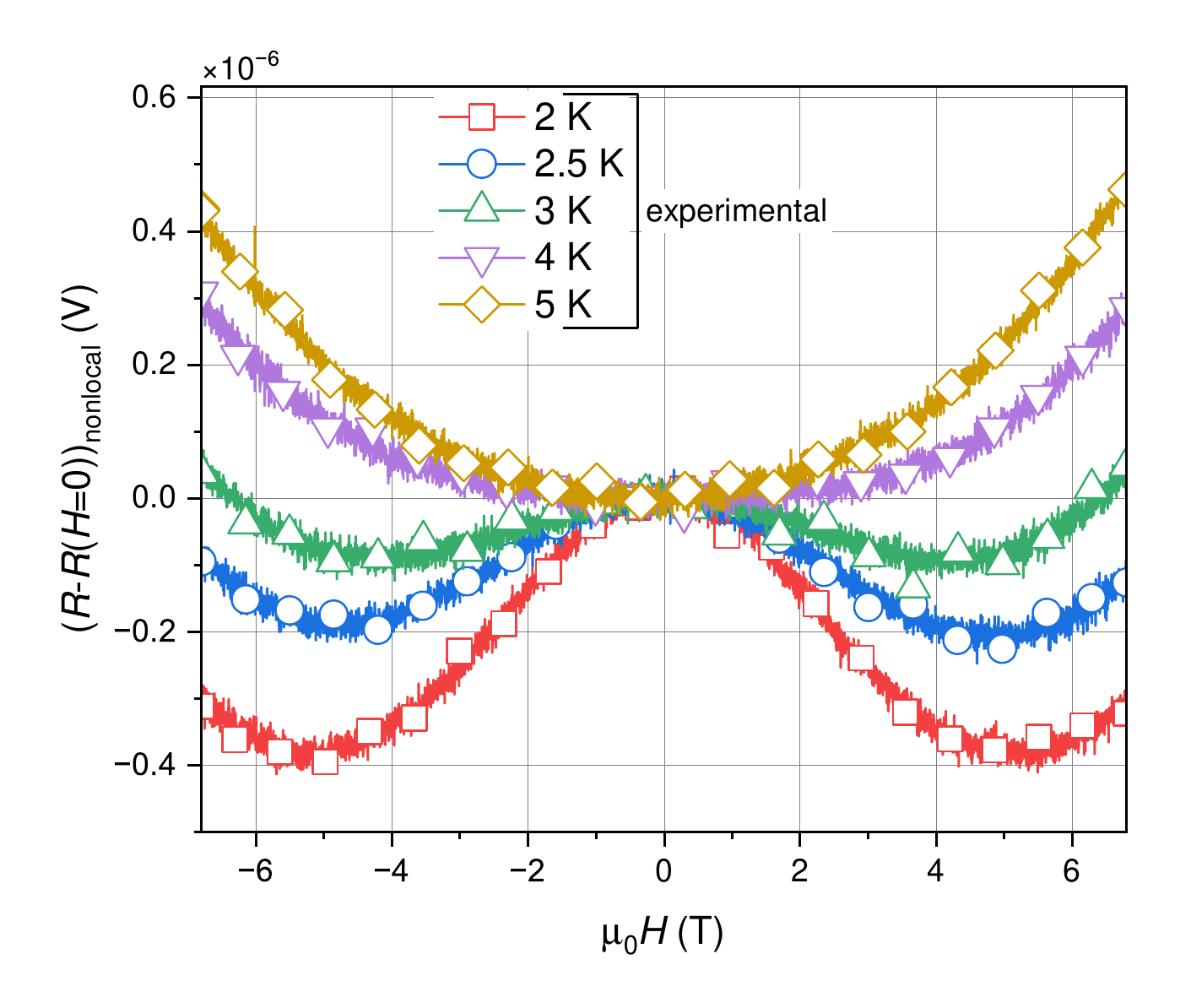}
\stackinset{c}{-0.8\linewidth}{c}{0.63\linewidth}{a)}{\hspace{0pt}}\\
\includegraphics[width=0.85\linewidth]{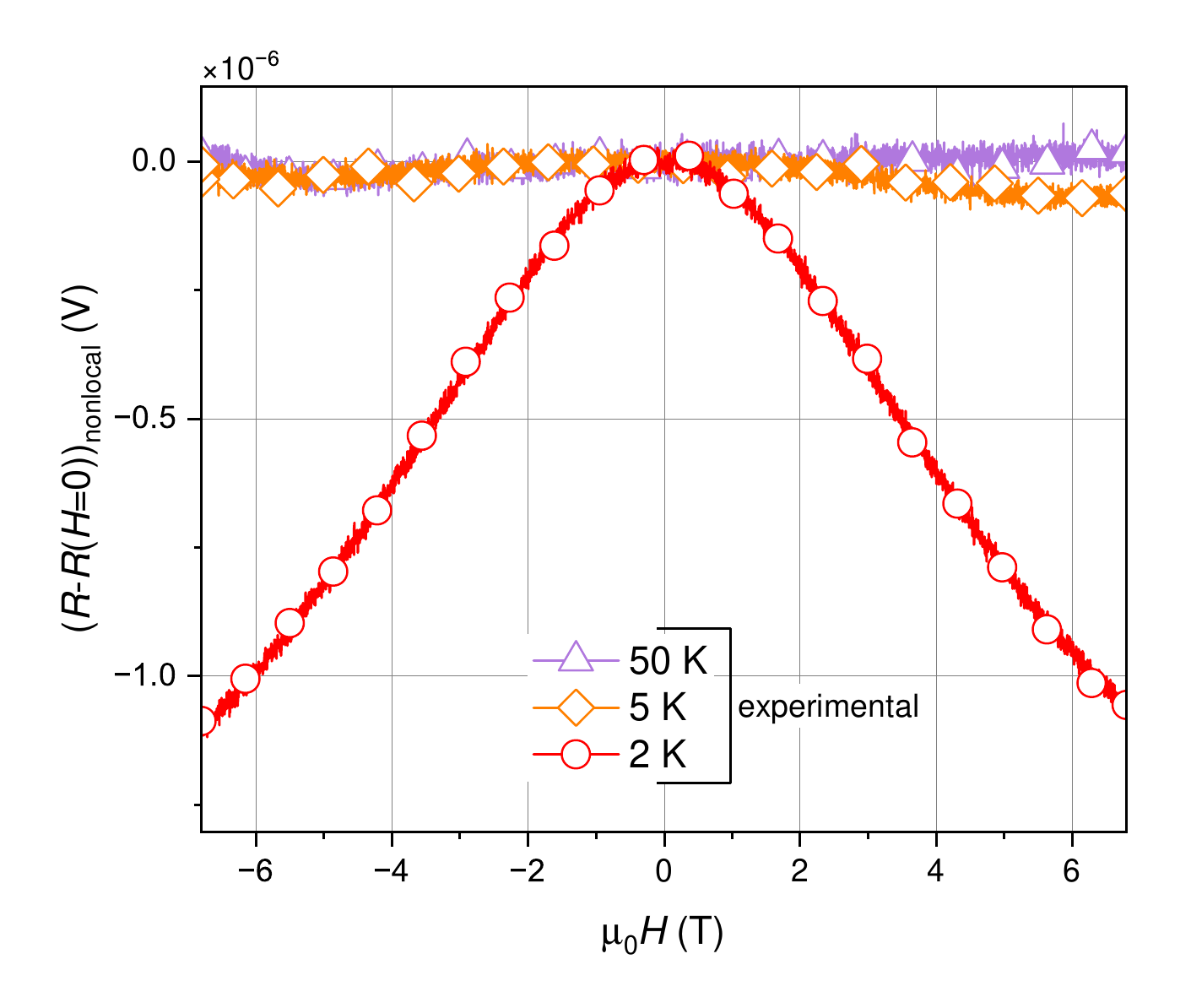}
\stackinset{c}{-0.8\linewidth}{c}{0.63\linewidth}{b)}{\hspace{0pt}}
\caption{Resistance $R-R(H=0)$ over applied magnetic field curves of sample A at various temperatures. a) For $(-90,-15)$ orientation. b) For $(-90,0)$ orientation.}
\label{i_MR_90_0and15_Ti_A}
\end{figure}

The stability of the \gls{nlmr} with respect to $T$ and $\psi$ at $\theta=\pm 90\,\degree$ ($\boldsymbol{H}$ rotating in the sample plane) is explored in order to discriminate the Kondo effect from a \gls{cme}: For sample B at $2\,\text{K}$, the \gls{nlmr} persists for arbitrary $\psi$, as depicted in \fig \ref{i_MR90_psi}, where the resistance $\left(R-R(H=0)\right)$ over $H$ is shown at specific values of $\psi$. Larger $\psi$-angles result in higher resistances towards high fields. In the orientation $(-90,15)$, sample A shows a resistance minimum for $\mu_0 H\approx \pm 5\,\text{T}$ at $2\,\text{K}$, as seen in \fig \ref{i_MR_90_0and15_Ti_A}\,a), depicting the resistance $\left(R-R(H=0)\right)$ over $H$ at specific temperatures. Upon increasing $T$, the \gls{nlmr} diminishes, transitioning to a purely positive \gls{mr} between $3\,\text{K}$ and $4\,\text{K}$.

The measurements elucidate the competition between the Kondo effect and the orbital effect: While the Kondo effect causes a \gls{nlmr} and is isotropic, the in-plane orbital effect scales approximately quadratically, but requires components $\boldsymbol{H} \bot \boldsymbol{E}$, which increase for $\psi\rightarrow\pm 90$. In sample B, even at $\psi = \pm 90$, a field of $\mu_0 H= \pm 6.8\,\text{T}$ is not sufficient to reach a regime dominated by the orbital effect. An orientation of $(-90,15)$ in sample A is employed to shift the minimum resistance below $6.8\,\text{T}$. The contribution from the Kondo effect diminishes with increasing temperature, resulting in the minimum resistance shifting to $H=0$ as the orbital effect dominates the system. No orbital effect emerges if the sample is rotated to $(\pm 90,0)$ and the resulting \gls{mr} is a constant for temperatures $T\geq 5\,\text{K}$, as seen in \fig \ref{i_MR_90_0and15_Ti_A}\,b), where the resistance $\left(R-R(H=0)\right)$ over $H$ is shown at specific temperatures.

\begin{figure}
    \includegraphics[width=0.93\linewidth]{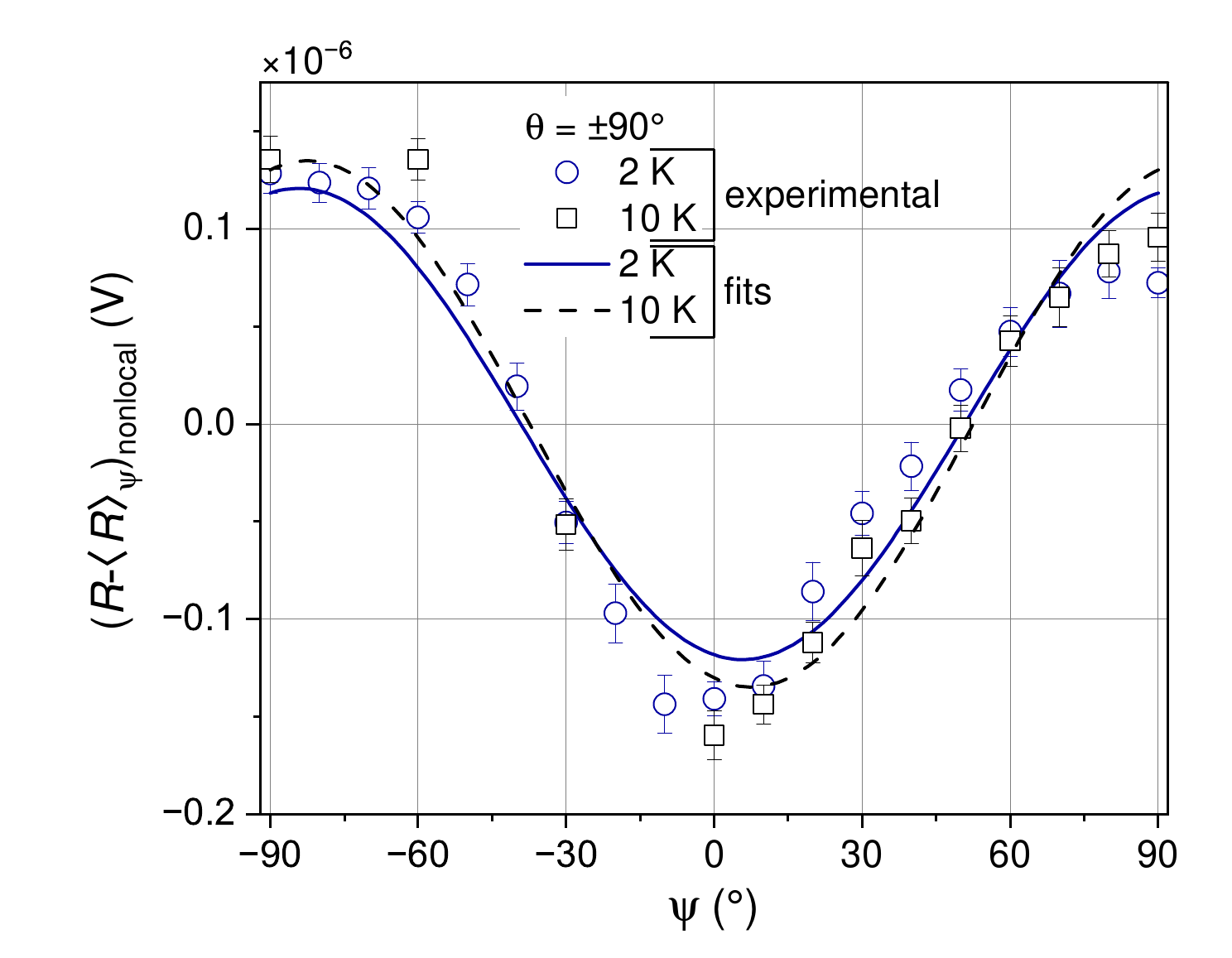}
    \stackinset{c}{-0.82\linewidth}{c}{0.63\linewidth}{a)}{\hspace{0pt}}\\
    \includegraphics[width=0.85\linewidth]{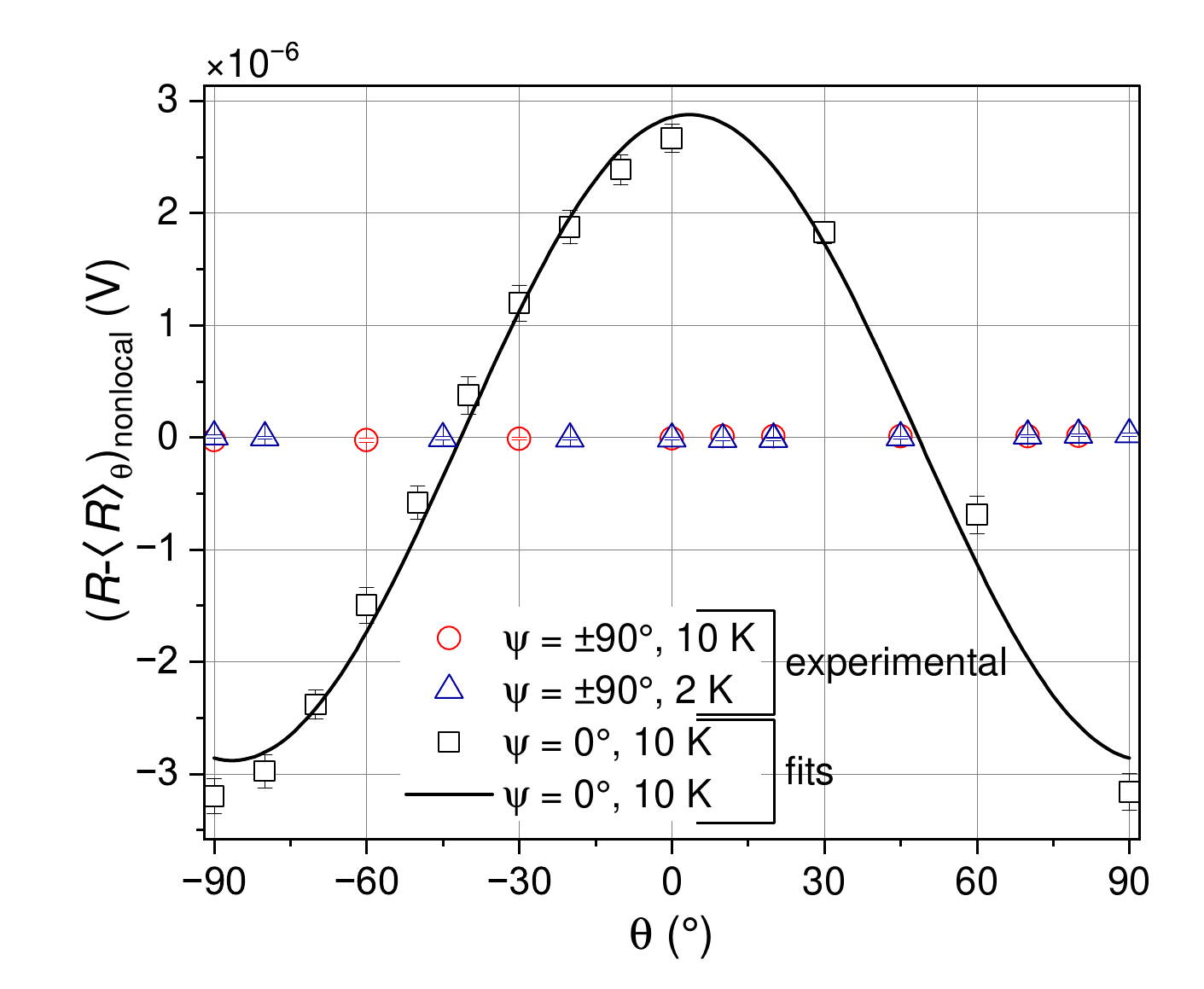}
    \stackinset{c}{-0.82\linewidth}{c}{0.6\linewidth}{b)}{\hspace{0pt}}
    \caption{Resistance upon rotating sample B at $6.8\,\text{T}$ for $2\,\text{K}$ and $10\,\text{K}$. a) Sweeping $\psi$ at $\theta = \pm 90\,\degree$. Continuous and dashed lines: Fits for $2\,\text{K}$ and $10\,\text{K}$, respectively. Resistance is given as deviation from the mean value of the fitting function. b) Sweeping $\theta$ at $\psi=0\,\degree$ and $\psi=\pm90\,\degree$. Line: Fit for $\psi=0\,\degree$ and $10\,\text{K}$. Resistance for $\psi=0\,\degree$ given as deviation from the mean value of the fitting function and for $\psi=\pm90\,\degree$ as deviation from the mean value.}
    \label{i_sweeps}
\end{figure}

This motivates a description of the resistance for arbitrary values of $T$,$H$,$\psi$ and $\theta$, which requires the consideration of the semimetallic term $R_\text{SM}(T)$, of the Kondo-term $R_\text{K}(H,T)$ and of the orbital terms $R_{\text{orb},\mathfrak{o}}(H)$, which are modulated by an angular dependence:
\begin{equation}
\begin{split}
R_{\text{orb},\perp}&\propto (\cos\theta)^2 (\cos\psi)^2,\\
R_{\text{orb},\parallel}&\propto (\text{const.}(\theta)) (\sin\psi)^2.
\end{split}
\end{equation}
To substantiate the angular dependence of $R_{\text{orb},\mathfrak{o}}$, the orientation of sample B is swept at a constant field of $H=6.8\,\text{T}$ while the longitudinal resistance is measured. The resistance $(R-\left\langle R \right\rangle_{\{\theta,\psi\}})$ as a function of $\psi$ or $\theta$ is shown in \fig \ref{i_sweeps}, with $\left\langle R\right\rangle_{\{\theta,\psi\}}$ being the average of $R$ over the respective angle $\theta$ or $\psi$. The in-plane-sweep results in a minimum resistance for $\psi=0\,\degree$, as depicted in \fig \ref{i_sweeps}\,a), where no (positive) orbital terms are present. The out-of-plane-sweep is shown for $\psi=0\,\degree$ in \fig \ref{i_sweeps}\,b), presenting the maximum resistance at $\theta=0\,\degree$, for which the maximum out-of-plane orbital magnitude is expected. The out-of-plane-sweep for $\psi=\pm 90\,\degree$, also reported in \fig \ref{i_sweeps}\,b), shows expectedly negligible effect, since the sweep corresponds to a rotation axis parallel to $\boldsymbol{H}$.

The general description for the longitudinal resistance is given by:
\begin{equation}
    R(H,T,\theta,\psi) = R_\text{SM}(T)+R_\text{K}(H,T)+\sum_\mathfrak{o} R_{\text{orb},\mathfrak{o}}(H,\theta,\psi).
    \label{eq_master_simple}
\end{equation}
The expanded form is provided in \eq \ref{eq_master} in the Supplemental Material.

The \gls{fc} measurements are carried out by cooling the sample in a constant magnetic field while measuring the longitudinal resistance. Using \eq \ref{eq_master}, \glspl{zfc} and \glspl{fc} can be fitted for arbitrary values of $\theta$ and $\psi$, as shown for sample B in the previously discussed \fig \ref{i_FCs}\,a). The magnetic field direction merely affects the amplitude of the orbital \glspl{mr}, which are only weakly temperature dependent for $T\lesssim 50\,\text{K}$. Therefore, \gls{fc} curves in different orientations differ mainly by a constant offset. In the range of $T < 10\,\text{K}$, the deviation between the fits and the data is a consequence of all curves sharing the parameters $T_\text{K} = 4.59\,\text{K}$ and $S = 0.086$. This suggests, that the description can be refined by employing the actual (unknown) magnetization $\boldsymbol{M}(\boldsymbol{H},T)$ of the vacancies instead of $\mathfrak{B}_J$.

\begin{figure}[htb]
\includegraphics[width=0.9\linewidth]{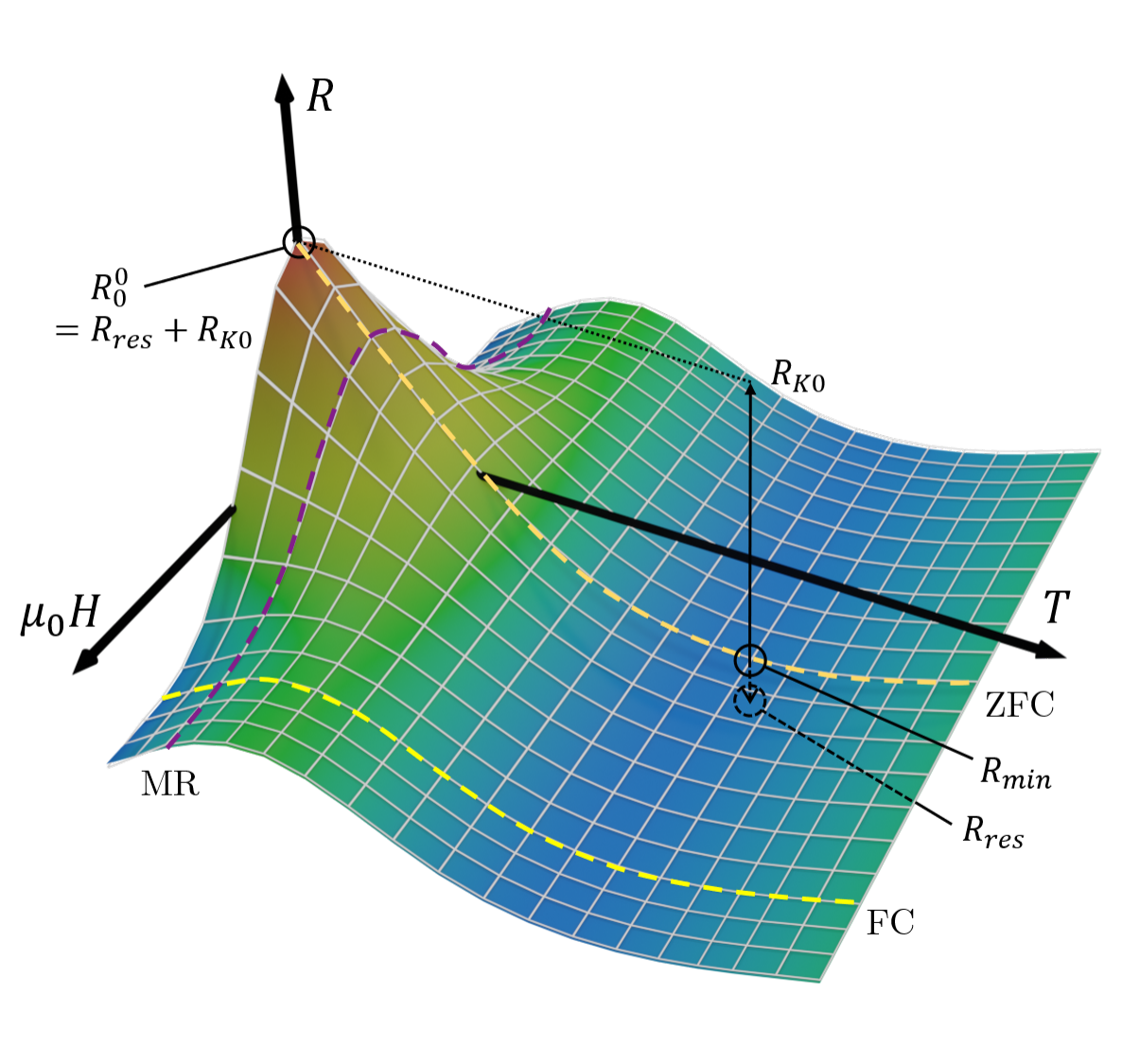}
\caption{Visualization of $R(H,T)$ as a function of temperature and magnetic field in the $(90,0)$ orientation. Relevant resistance points defining the shape of the surface are marked. The point $R_\text{res}$ corresponds to the minimum resistance without a Kondo effect and lies below the surface. Plot not to scale.}
\label{i_KondoSurf}
\end{figure}

A visual representation of the magnitude of the Kondo effect is provided in \fig \ref{i_KondoSurf}: the rendered $R(H,T,\theta=90,\psi=0)$-surface represents the $T$- and $H$-dependencies and includes relevant resistance points that define the shape of the surface in parameter space.

The in-plane behavior observed in \fig \ref{i_MR90_psi} is also fitted satisfactorily. However, for $\mu_0 H \gtrsim 5\,\text{T}$, it can be gleaned that the curve produced by the fit slightly underestimates the resistance for small $\psi$ and somewhat overestimates it for $\psi\rightarrow \pm 90\,\degree$. This again points to a deviation between $\boldsymbol{M}(\boldsymbol{H})$ and $\mathfrak{B}_J(H,T)$

\subsection{Carrier density}
\begin{figure}[htb]
\includegraphics[width=0.85\linewidth]{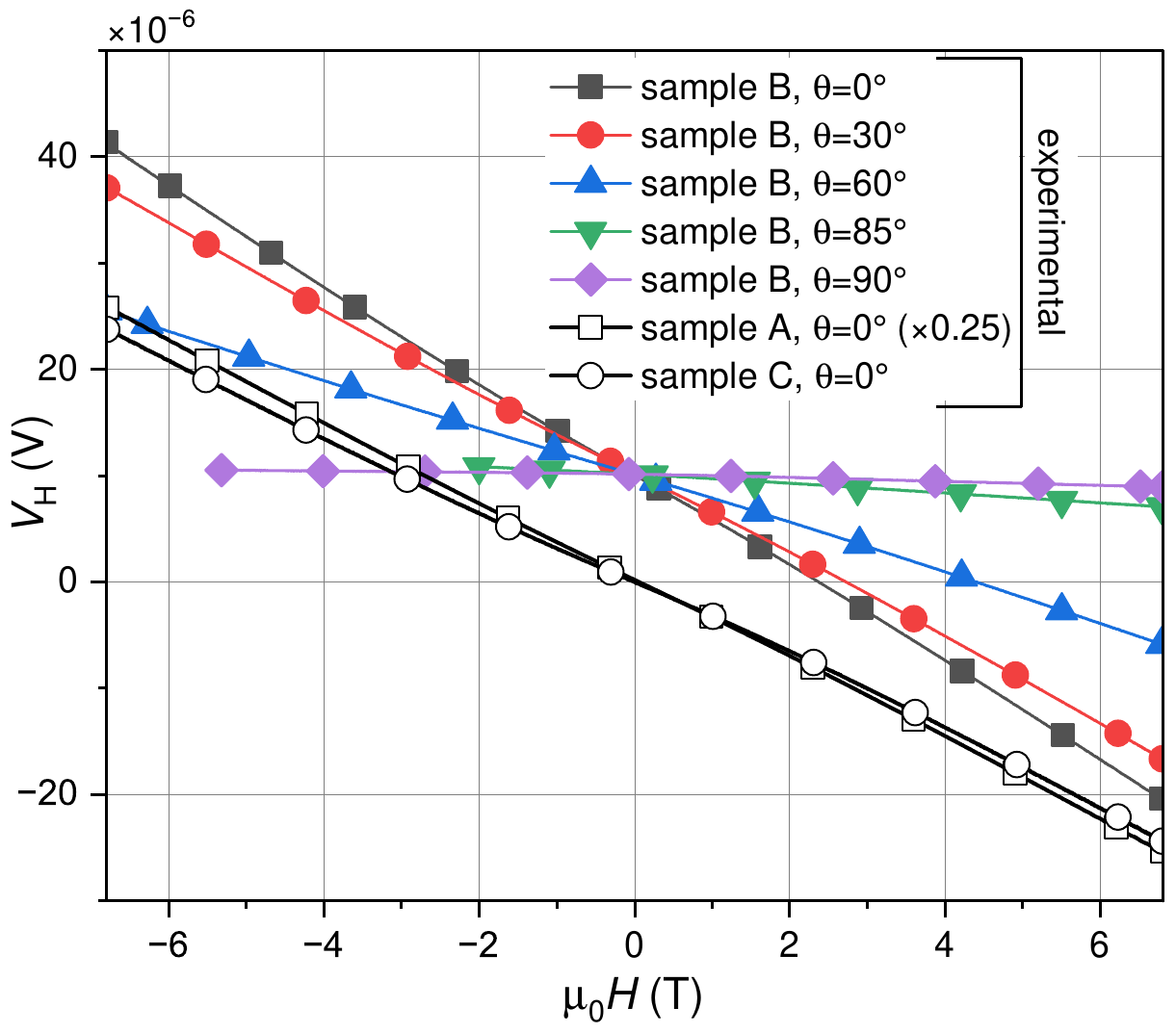}
\caption{Hall voltages $V_H$ over applied magnetic field at $2\,\text{K}$ in the $(0,0)$ orientation for samples A, B and C. For sample B, $\theta$ is varied from $(0,0)$ to $(90,0)$ . Hall voltage of sample A at $1/4$ scale.}
\label{i_Hall_2K}
\end{figure}

For terminals oriented normal to the drain-source direction, a Hall-voltage is obtained. Measurements thereof at $2\,\text{K}$ in the $(0,0)$ orientation for samples A, B and C are reported in \fig \ref{i_Hall_2K}, showing the Hall voltage $V_H$ as a function of the applied magnetic field: A linear dependence of $V_H \propto H$ is observed. As sample B is tilted from $(0,0)$ towards $(90,0)$, the out-of-plane component of $\boldsymbol{H}$ is reduced, which in turn reduces the slope of the Hall voltage.

The Hall curves of \fig \ref{i_Hall_2K} yield at $2\,\text{K}$ charge carrier concentrations $n=\frac{I \mu_0 H}{V_H t q}$ of $(2.33; 7.02; 6.93) \times 10^{20}\,\text{cm$^{-3}$}$ for samples A, B and C, respectively. Here, the applied current is $I = 10\,\mu\text{A}$ and $q$ is the elementary charge. In sample B, the constant offset of $\sim 10\,\text{$\mu$V}$ for zero field does not diminish with $\theta$ and is therefore attributed to an asymmetric sample geometry resulting in a longitudinal transport component and, thus, an apparent $V_H$ at zero field.

\begin{figure}[htb]
\includegraphics[width=0.95\linewidth]{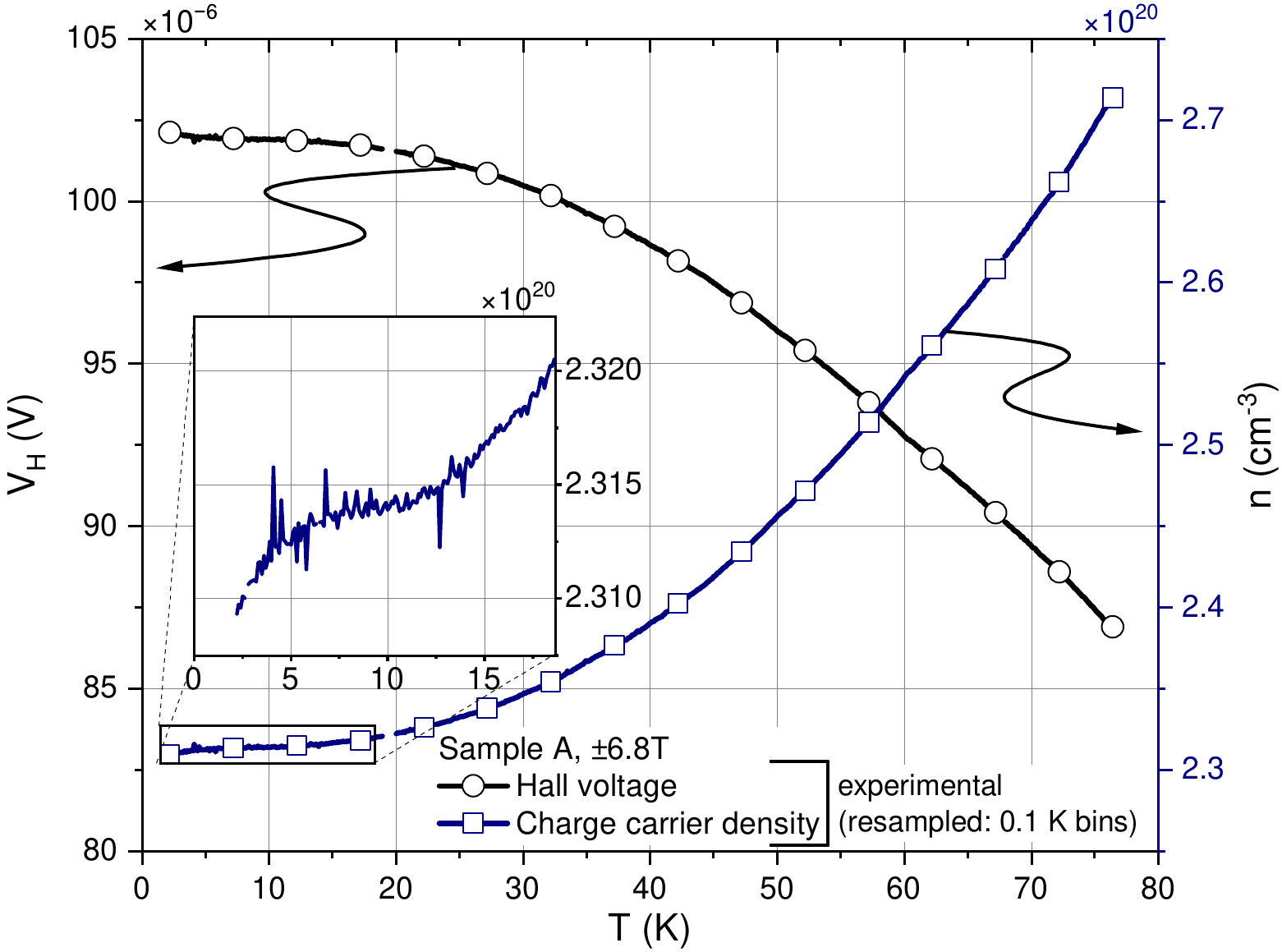}
\caption{Computed Hall voltage $V_H$ and charge carrier density $n$ over temperature for sample A. Inset: $n$ for $T<18\,\text{K}$.}
\label{i_Hall_FC}
\end{figure}

For sample A, \glspl{fc} in Hall-configuration are obtained at $\mu_0 H=\pm 6.8\,\text{T}$. The Hall voltage $V_H$ computed according to:
\begin{equation}
V_H(H,T)=\frac{1}{2}\left(V(H,T)-V(-H,T)\right),
\label{eq_Hall}
\end{equation}
is plotted in \fig \ref{i_Hall_FC} together with the carrier density $n$ derived thereof.

The carrier density decreases with temperature, plateauing at $T \sim 10\,\text{K}$. For $T\rightarrow 0$, $n$ drops further, which can be understood in the frame of the strong-coupling Kondo regime wherein the impurity spins pair with charge carrier spins, effectively capturing them. This points to $T_\text{K}\approx 10\,\text{K}$, consistent with the values obtained from the $\text{ZFC}$ and $\text{FC}$ curves. An extrapolation to higher temperatures yields a charge carrier density ratio $\frac{n(300\,\text{K})}{n(2\,\text{K})} \approx 9$.

\FloatBarrier
\section{Modelling}
To substantiate the findings, \gls{dft} based calculations for Pt-surface-defects within \pt are carrier out. The investigated specimens thereby correspond to a multilayered \gls{2d} system. Notably, \pt exhibits an indirect bandgap exclusively in ultrathin layers, transitioning to a semimetallic state for a thickness $\geq$ three monolayers.
\begin{figure}[htb]
    \centering
    \includegraphics[width=0.95\linewidth]{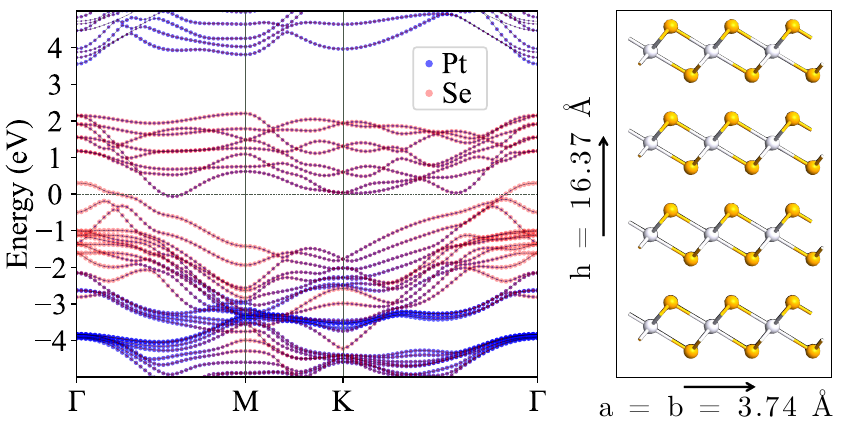}
    \stackinset{c}{-0.93\linewidth}{c}{0.48\linewidth}{a)}{ }
    \stackinset{c}{-0.35\linewidth}{c}{0.48\linewidth}{b)}{ }
    \caption{a) Bandstructure of four layers of \pt with contributions of Pt (blue) and Se (red) atoms. b) Calculated parameters for 4-layer unit cell of \pt.}
    \label{fig:bands4l}
\end{figure}
In the semimetallic phase, \pt exhibits a valence band maximum located at the $\mathit{\Gamma}$ point, while the conduction band minimum is found between the $M$ and $\mathit{\Gamma}$ points, as comprehensively demonstrated across layers-numbers ranging from four to 10 \cite{Villaos2019} and depicted in \fig \ref{fig:bands4l}\,a). A 4-layer model, sketched in \fig \ref{fig:bands4l}\,b) is considered. The computed Se-Se distance in the 4-layer unit cell is $h=16.37\,\text{Å}$, with a Pt-atom-layer positioned between two Se atom layers, forming a stable 1T phase characterized by a lattice parameter of $3.74\,\text{Å}$. The presence of Pt-defects in the samples was observed experimentally in \refText \cite{Avsar2019}. Notably, it has been established, that the formation of defects within the bulk does not result in the generation of a local magnetic moment \cite{Avsar2019}. This permits to focus the computation on the Pt-surface-defects within the 4-layer model. The inclusion of the Pt-vacancy defect at the surface of the 4-layer model results in the generation of a magnetic moment, averaging around $1.3\,\mu_B$. 

\begin{figure}[htb]
\includegraphics[width=0.85\linewidth]{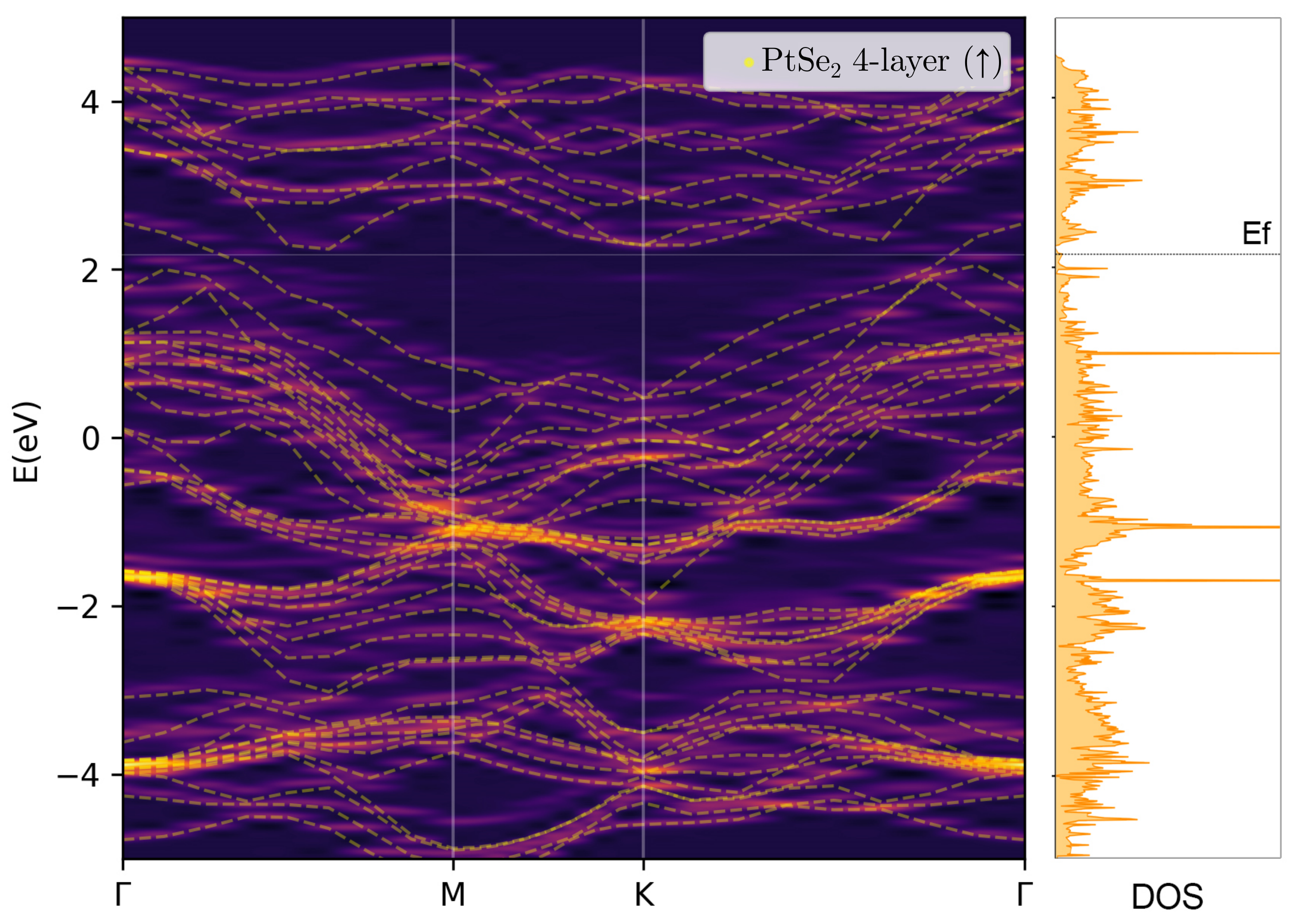}\\
\includegraphics[width=0.85\linewidth]{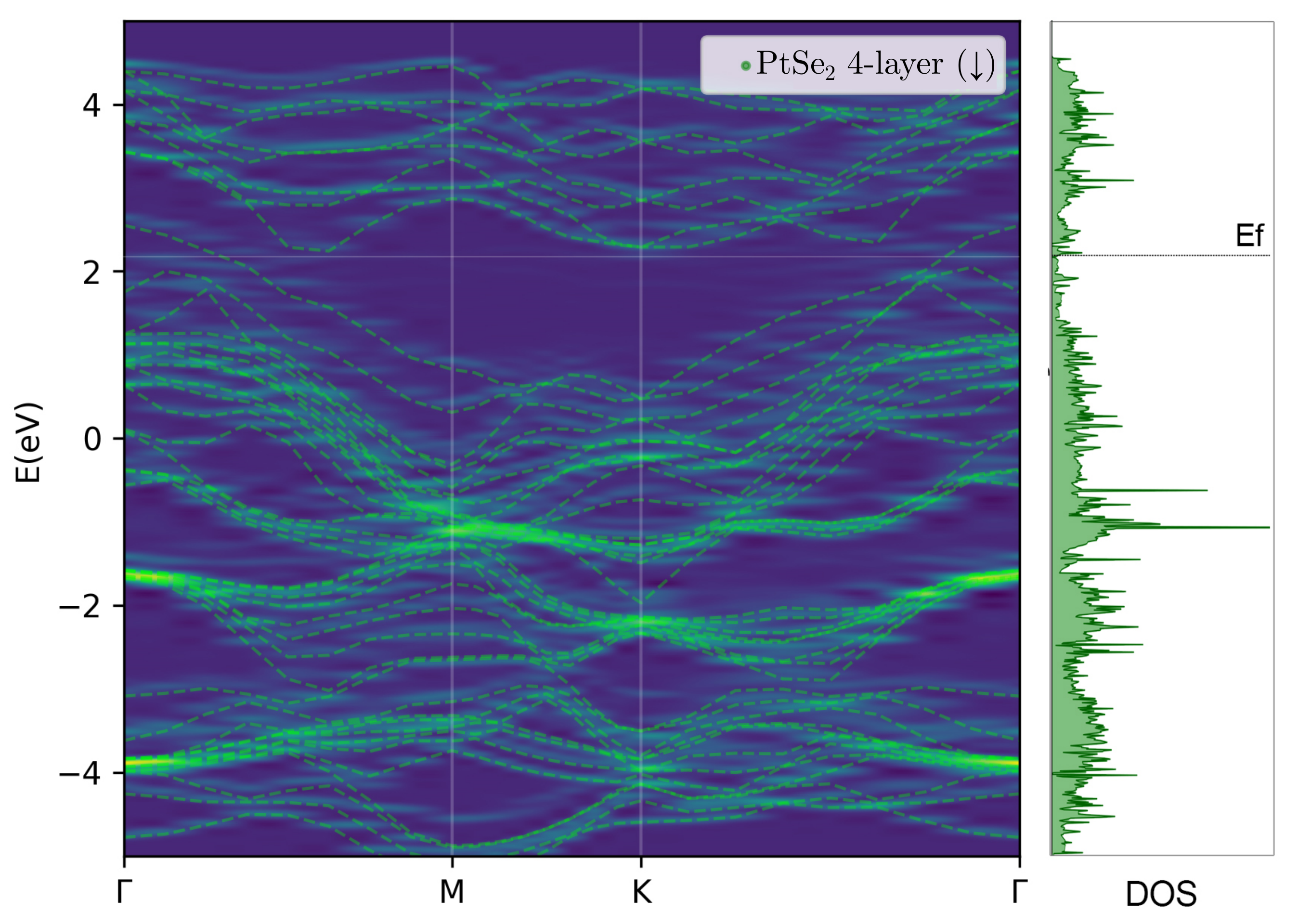}
\caption{Electronic \gls{dos} in the \pt 4-layer model with Pt-vacancy. Upper panel: Spin-up \gls{dos} map over momentum and energy (left) and spin-up \gls{dos} over energy (right). Bottom panel: Equivalent plots for spin-down \gls{dos}.}
\label{fig:vacBands}
\end{figure}

\begin{figure}[htb]
    \begin{tikzpicture}
        \node (image) at (0,0) {\includegraphics[width=0.95\linewidth]{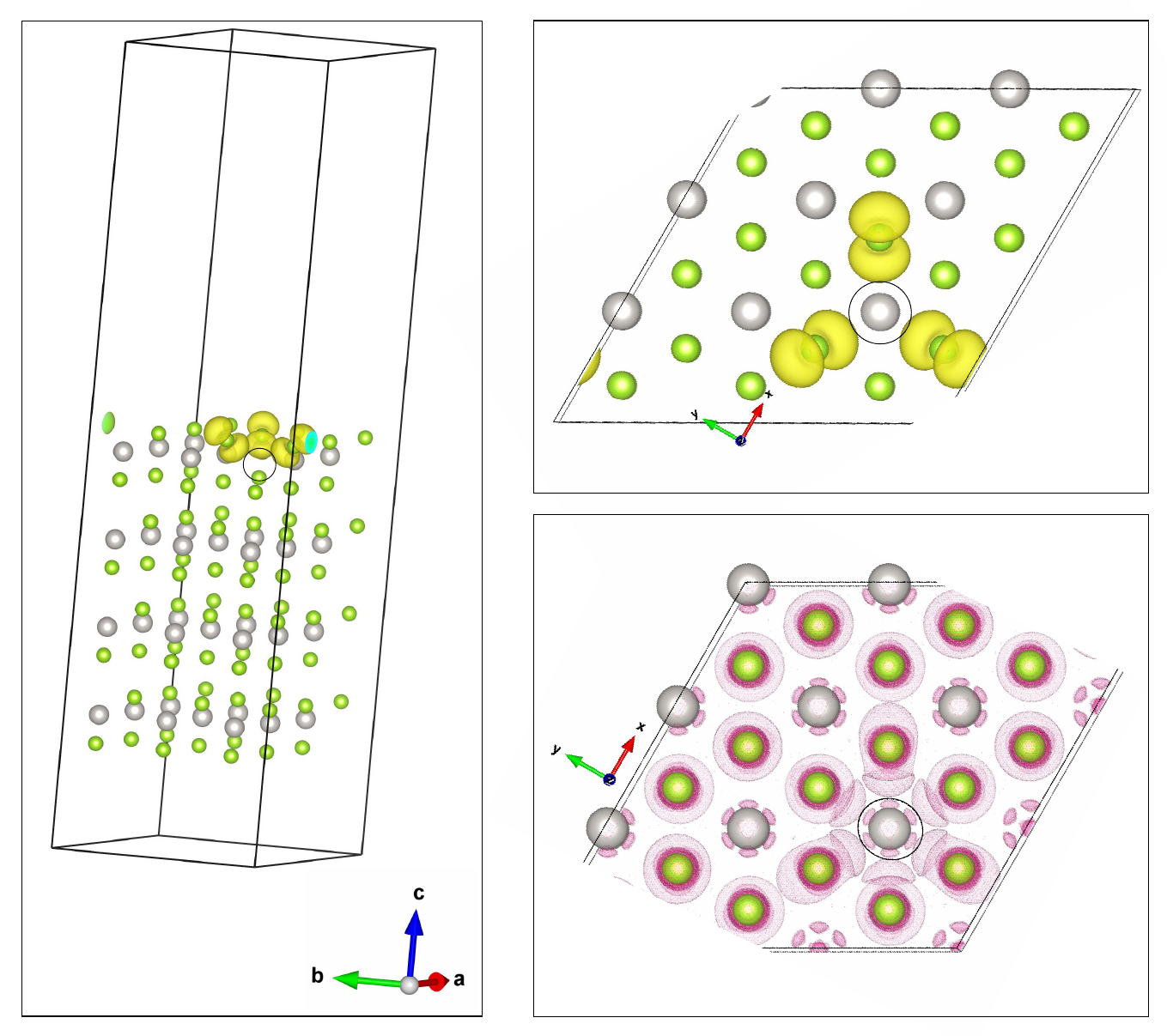}};
        \node at (-0.42\linewidth, 0.37\linewidth) {a)};
        \node at (-0\linewidth, 0.37\linewidth) {b)};
        \node at (-0\linewidth, -0.03\linewidth) {c)};
    \end{tikzpicture}
    \caption{Spin density (yellow) distribution around the Pt-vacancy (circle) in a multilayer model, viewed a) isometrically and b) along the out-of-plane axis. c) Charge density (brown) distribution around the Pt-vacancy (circle).}
    \label{fig:dens3D}
\end{figure}
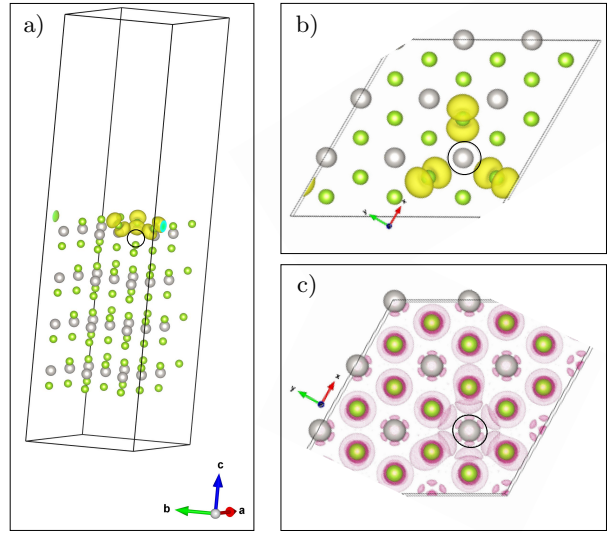

The electronic \gls{dos} shown in \fig \ref{fig:vacBands} is given for spin-up states (upper panels) and spin-down (lower panels). The analysis of the density-maps in momentum space (right panels) and the respective \gls{dos} over energy (left panels) points at a local magnetic moment originating from defect-induced spin-split states that emerge in proximity of the Fermi level (dashed line) \cite{Avsar2019}. The spin density surrounding the defect predominantly concentrates on the adjacent Se atoms within their immediate coordination sphere, as sketched in\fig \ref{fig:dens3D}\,a),b). Additionally, the Pt-vacancy is accompanied by a noticeable localization of charge density surrounding it, as shown in \fig \ref{fig:dens3D}\,c).

\section{Conclusion}
The electronic structure, and in particular the mechanism leading of the observed \gls{nlmr} for exfoliated \pt flakes are examined. Low $T$/high $\mu_{0}H$ magnetotransport is performed, yielding insights \textit{via} application of analytical fitting functions. The magnetotransport allows for the identification of a Kondo effect being present in all considered samples: The observed zero-field behavior can be understood using an analytical formalism, while the field-dependent effects are separated into an isotropic \gls{nlmr} stemming from the Kondo effect and angle-dependent conventional orbital \gls{mr} contributions. The attribution of the \gls{nlmr} to a \gls{cme} or to current jetting is discussed, resulting in their confutation by considering the angle-dependencies and the robustness of the observed \gls{nlmr}. A transport model is formulated to describe the total longitudinal \gls{mr} in \pt at arbitrary temperature, magnetic field and magnetic field direction. The model captures, how the Kondo effect emerges at temperatures $T\lesssim 15\,\text{K}$ and how the polarization of the impurity spins reduces Kondo scattering when an external magnetic field is applied. By using \gls{dft} in conjunction with transport performed on different flake thicknesses, the origin of the impurity spins is identified in Pt-defects, which contribute an uncompensated spin exclusively at the sample surface. It remains to derive and study the rigorous magnetization function $\boldsymbol{M}(\boldsymbol{H})$ of the magnetic moments. The results point at a \gls{2d} spin-density at the surfaces of \pt flakes and highlight both the sample thickness and the vacancy concentration as tuning parameters for interplay between the Kondo effect and orbital \gls{mr}. Understanding the dynamics of both the spin- and orbital degree of freedom of the charge carriers in \pt is fundamental to spintronic applications and detection of an orbital Hall effect.

\section*{Acknowledgements}
This work was funded by the Austrian Science Fund (FWF) through Project No. TAI-817 and by the JKU LIT Seed funding through Project No. LIT-2022-11-SEE-119.

\bibliography{bib}
\clearpage

\appendix
\begin{widetext}
\figureS
\tableS
\equationS
\section{Supplemental material}
\subsection{Fabrication}
The exfoliation of the \pt specimens is carried out from a HQ Graphene \cite{HQGraphene} bulk crystal using Nitto tape type ELP BT-150P-LC. Between four-to-six exfoliation steps result in an acceptable yield of thin flakes. Pressure is applied to the Nitto tape/\pt/Nitto tape stack with fingers from both sides, followed by a rapid peel off. The last exfoliation step is carried out with a Nitto tape/\pt/\gls{pdms} stack. The \gls{pdms} itself is mounted onto a stiff polymer for controlled movement during the deterministic dry transfer, which allows to precisely place the selected flake onto the contacts. The maximum setting of the dry transfer illuminator results in a sample temperature of $\approx 45\,\degree\text{C}$, easing the lift-off for the flakes away from the \gls{pdms}. The substrate is heated to $110\,\degree\text{C}$ (\textit{via} a hot plate) to avoid water-adsorbate reducing the \gls{vdw}-bond to the flake. The contacts feature a Hall-bar design with dedicated source and drain contacts. The contact wires end in $100\,\mu\text{m}$ diameter contact pads. Electron beam lithography is used to pattern the contacts onto the $90\,\text{nm SiO}_2/ \text{Si}\left\langle 100 \right\rangle p^{++}$ substrates. The resist is CSAR 9200.09 spin coated at $4\,\text{krpm}$ on clean substrates preheated to $150\,\degree\text{C}$, then soft baked on a hot plate at $150\,\degree\text{C}$ for two minutes. The structure is pattered with a charge dose of $18\,\mu\text{C/cm}^2$ at $10\,\text{kV}$ extra-high tension and a $20\,\text{$\mu$m}$ aperture. The development of the resist is performed by ARP 600-549 for one minute and stopped by \gls{ipa}. A Pt-film of $10\,\text{nm}$ in thickness is sputtered using Argon plasma at a pressure of $10^{-3}\,\text{mbar}$ and a current of $25\,\text{mA}$. Lift-off is carried out by anisole at $65\,\degree\text{C}$ on a hot plate for 15 minutes, followed by rinsing using a pipette to wash the sample with the selfsame hot anisole. Final cleaning is performed \textit{via} \gls{ipa} and nitrogen gun. The contact pads are wire-bonded with $25\,\mu\text{m}$ gold wire fed through a ceramic tip bonder. The solder agent is In at $200\,\degree\text{C}$. Devices thus fabricated show two-terminal resistances in the $(5-10)\,\text{k}\Omega$ range. The flake selection is based on optically estimated thickness, size, shape and surface smoothness, as well as thickness uniformity.

\subsection{Atomic force microscopy}
Atomic force microscopy is employed to determine the thickness of the \pt flakes and the Pt-contacts. The cantilever is operated in tapping mode. An atomic force microscopy image of sample A is provided in \fig \ref{i_AFM}, showing sample height as a heatmap. The fact that the flake edges and steps edges meet at $120\,\degree$- and $60\,\degree$-angles indicates a $C_3$ or $C_6$ in-plane symmetry and high crystallinity.

\begin{figure}[htb]
\includegraphics[width=0.5\linewidth]{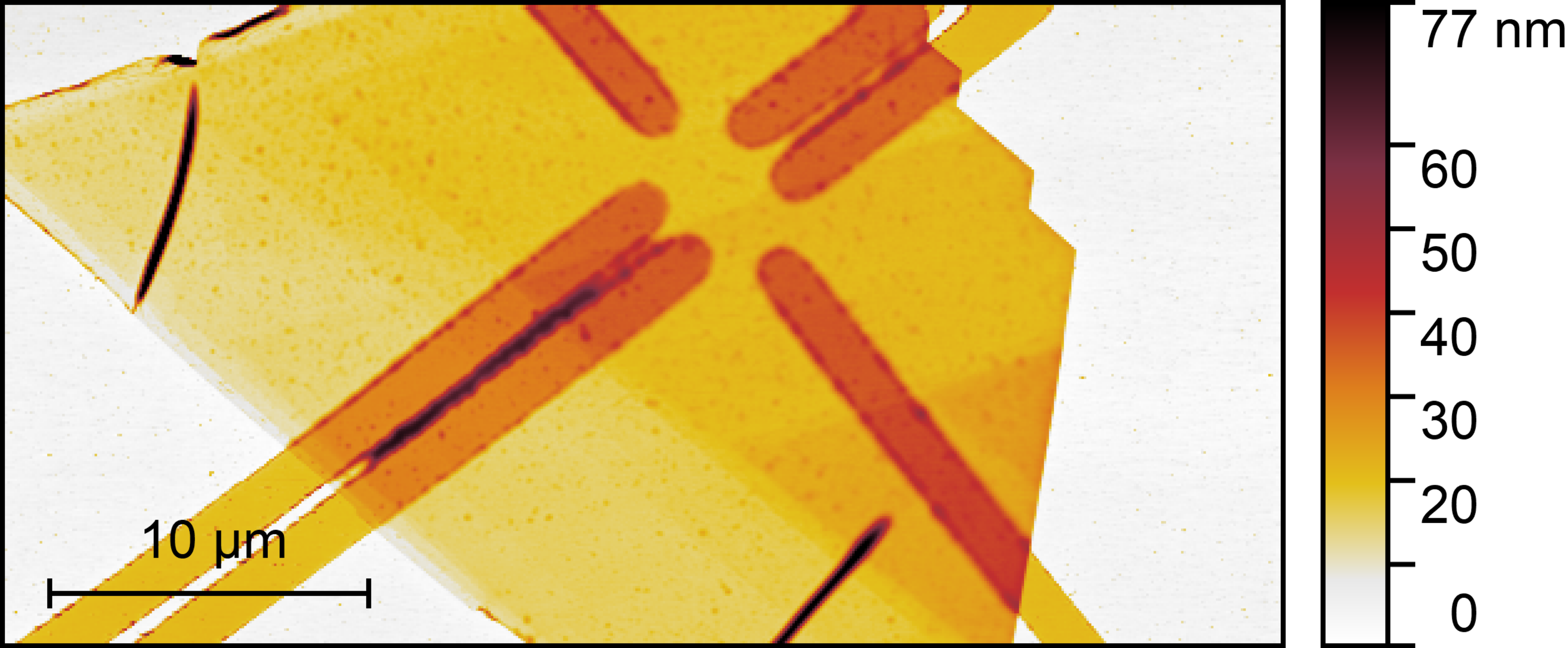}
\caption{Atomic force microscopy image of sample A. Heatmap of sample height over surface position.}
\label{i_AFM}
\end{figure}

\FloatBarrier
\subsection{Raman spectroscopy}
Raman spectroscopy is performed on the \pt flakes using a $532\,\text{nm}$ laser at $1.5\,\text{mW}$ with $\sim 1\,\mu\text{m}$ spot size. The integration time is $100\,\text{s}$. \fig \ref{i_Raman} shows the spectrum of intensity wavenumber-shift of both a bulk flake (thickness in the order of $100\,\text{nm}$) and a thin flake (thickness in the order of $10\,\text{nm}$). The observed $E_g$- and $A_{1g}$-peaks are in agreement with results from Mingzhe Yan \textit{et al.} \cite{MingzheRaman}, considering the difference in laser wavelength ($633\,\text{nm}$). The increased visibility of the $LO$ phonon and the presence of a substrate peak are expected for a thin \pt layer. The qualitative features identified allow the conclusion that the flakes are of space group \#164. The peak positions support an identification as \pt.

\begin{figure}[htb]
\includegraphics[width=0.4\linewidth]{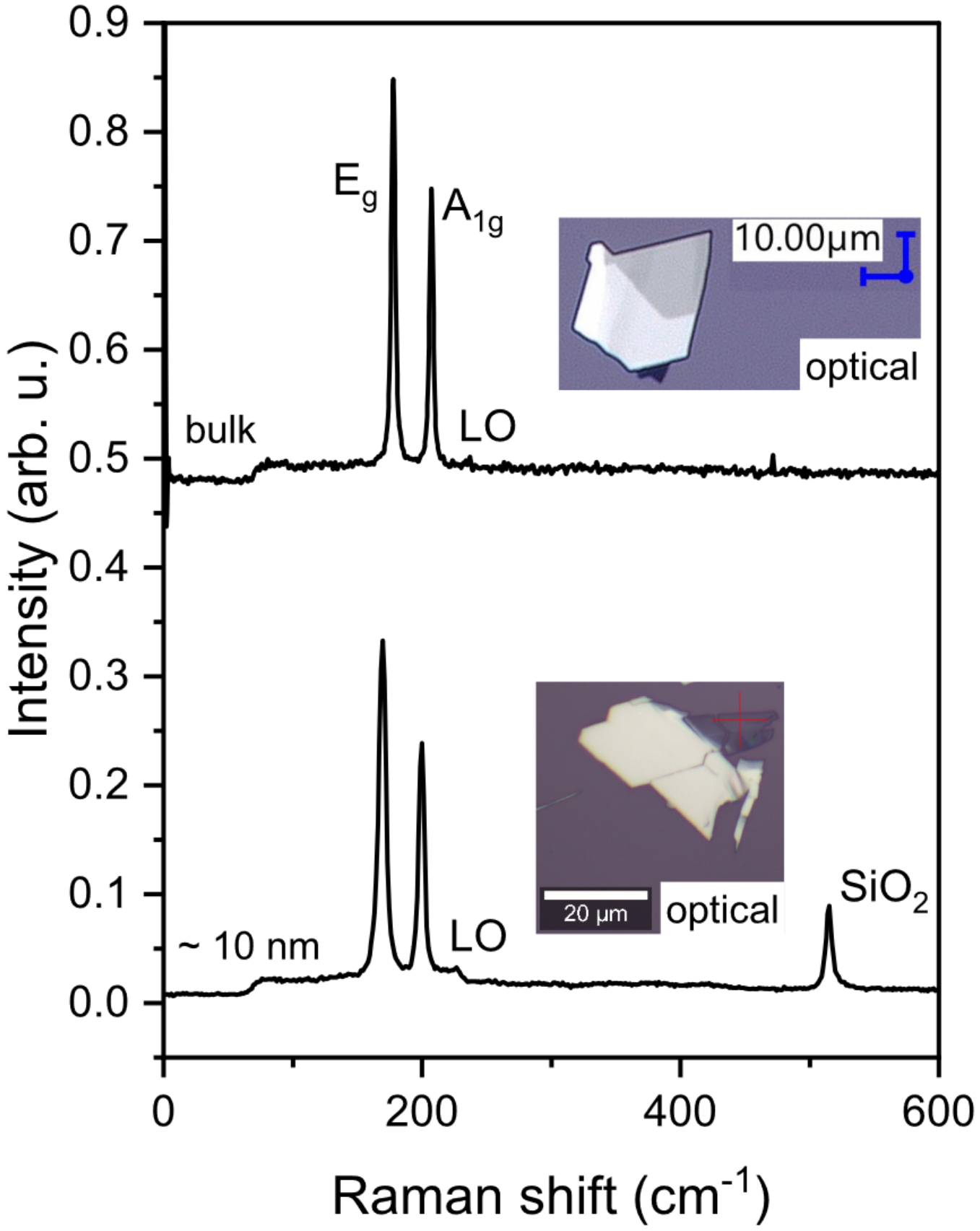}
\caption{Intensity over wavenumber-shift of \pt flakes on $\text{SiO}_2$. Bulk spectrum (top) and $\sim 10\,\text{nm}$ (bottom) with insets showing the respective optical images.}
\label{i_Raman}
\end{figure}

\subsection{Sample stability in ambient atmosphere}
\gls{xps} data is acquired on a series of \pt samples after increasing exposure time to ambient atmosphere. \fig \ref{i_XPS_air} shows the \gls{xps} spectra, counts over binding energy, over time. This confirms the presence of Se and Pt in the flakes and agrees with other works on \pt \gls{xps} \cite{MarinovaPtSe2char}\cite{DawsPtSe2char}. The \gls{xps} results support the Raman data, and the flakes can be confidently identified as \pt. The samples have been kept in dry conditions in ambient atmosphere for up to six months, which has resulted in a significant shift of the peak energies towards higher binding energies $E_B$. This suggests oxidation taking place over the time-frame of days for Se and months for Pt. However, there is little qualitative change in the shape of the Pt- and Se-peaks, - additional oxidation peaks being expected. Alternatively, the shift could stem from potentially ambiguous binding energy corrections during analysis due to different C species being present on the different samples following the prolonged ambient atmosphere exposure. In \fig \ref{i_XPS_air_fit}, counts over binding energy for samples exposed three days to ambient atmosphere are shown, demonstrating that the spin-split peaks are a reasonable fit to the acquired data.

\begin{figure}
\includegraphics[width=0.4\linewidth]{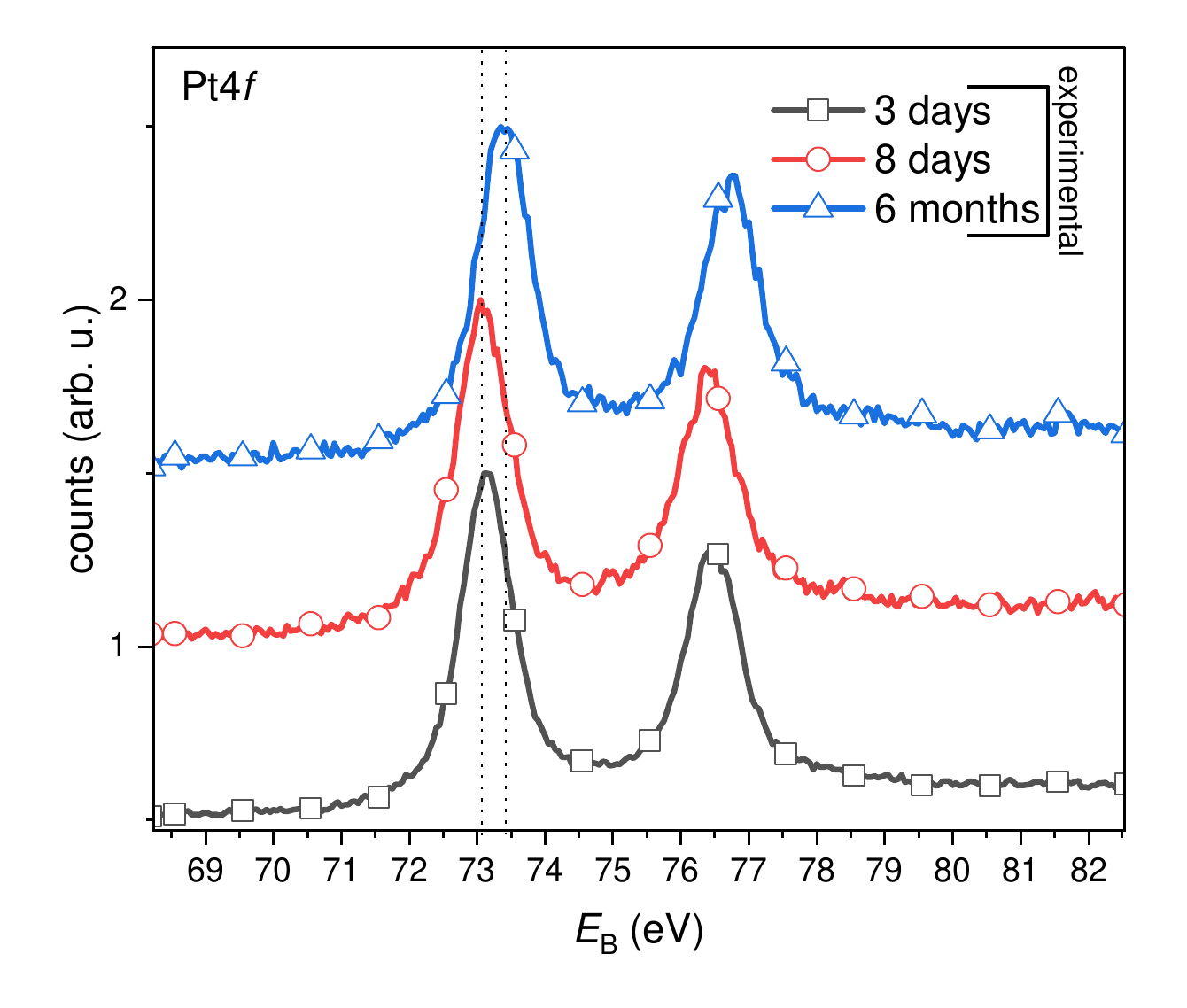}
\includegraphics[width=0.4\linewidth]{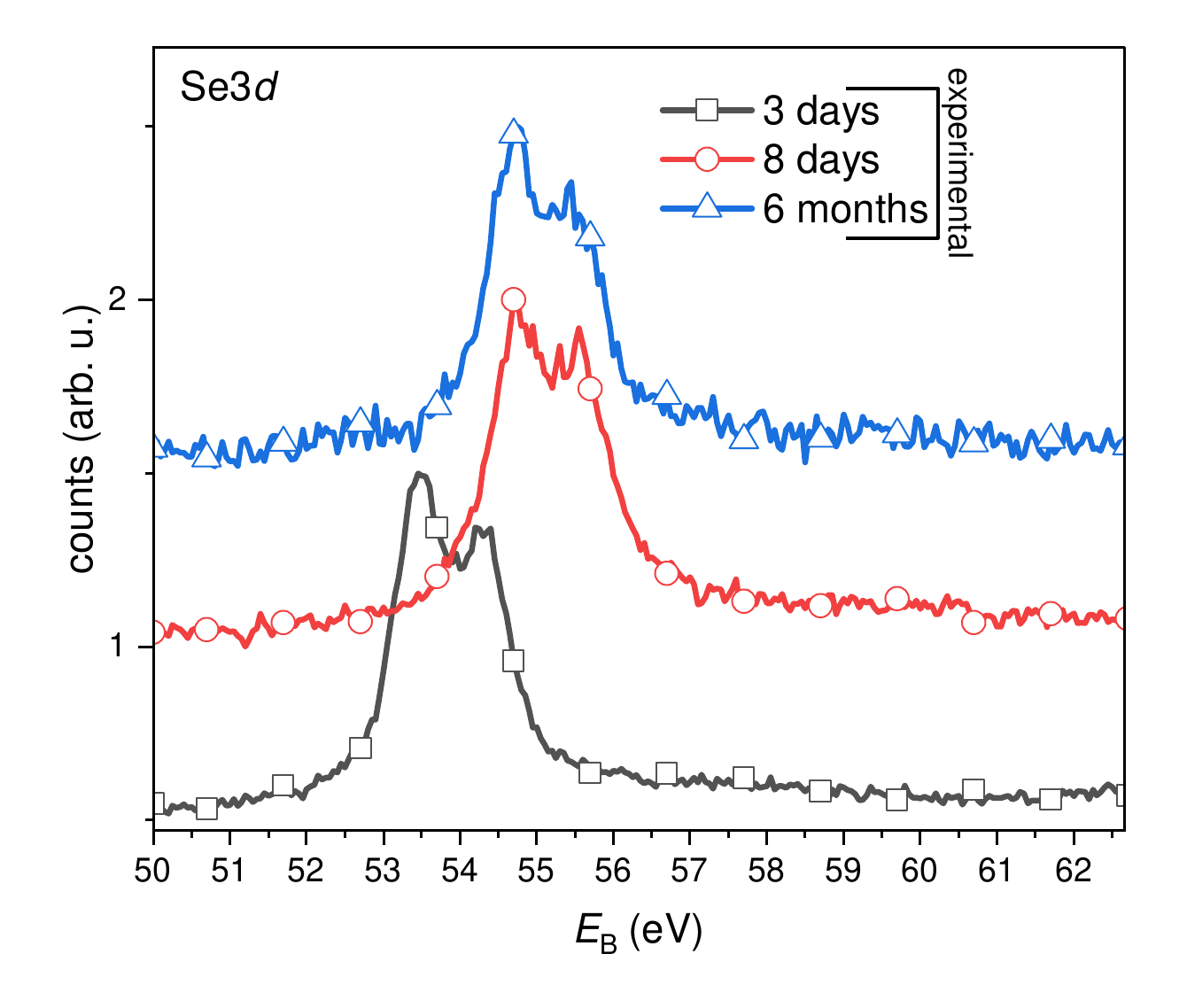}
\caption{XPS spectra, counts over binding energy, of \pt flakes after ambient air exposure for up to six months. Shown are the Pt4$f$ spectrum (right) and Se3$d$ (left).}
\label{i_XPS_air}
\end{figure}

\begin{figure}
\includegraphics[width=0.4\linewidth]{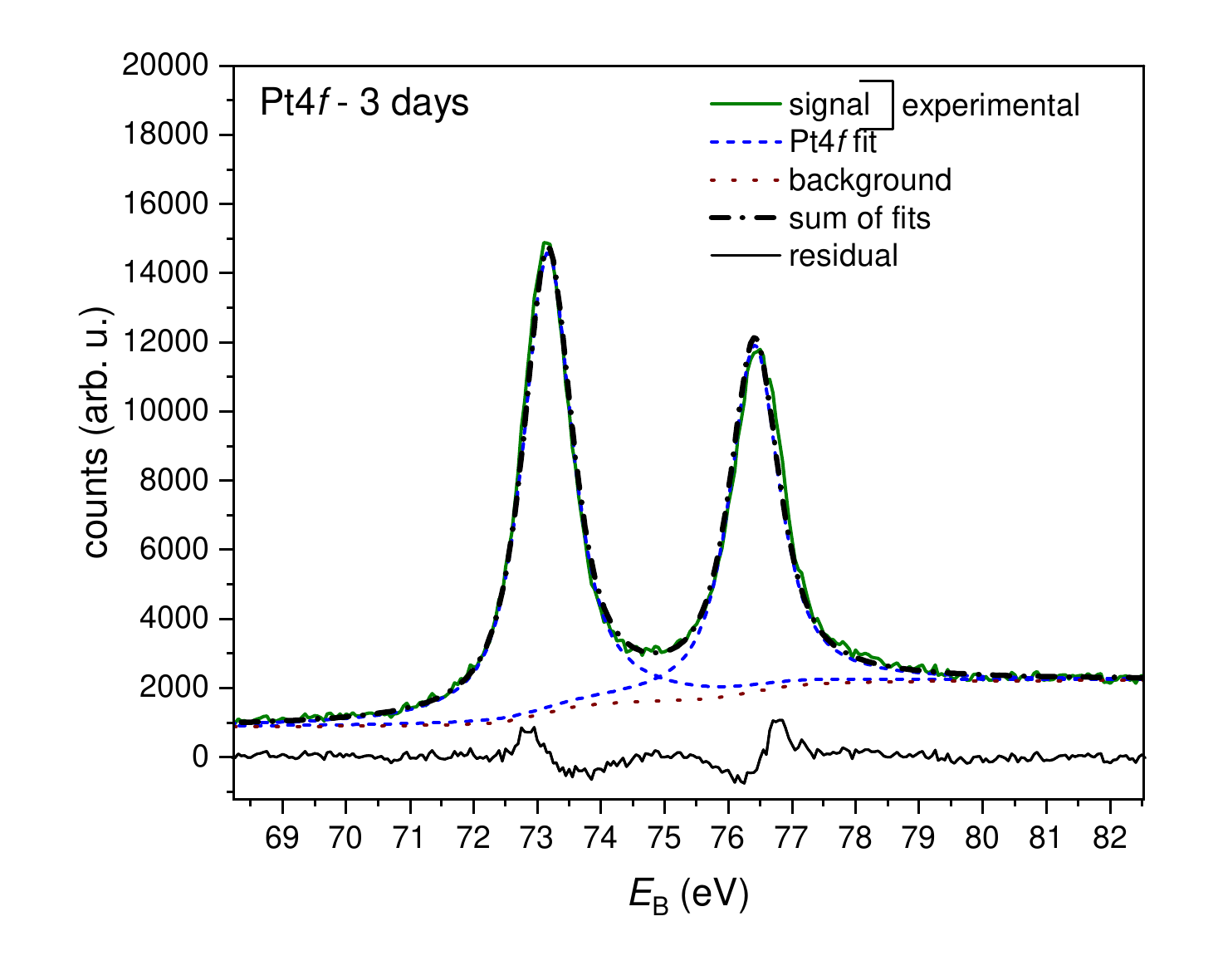}
\includegraphics[width=0.4\linewidth]{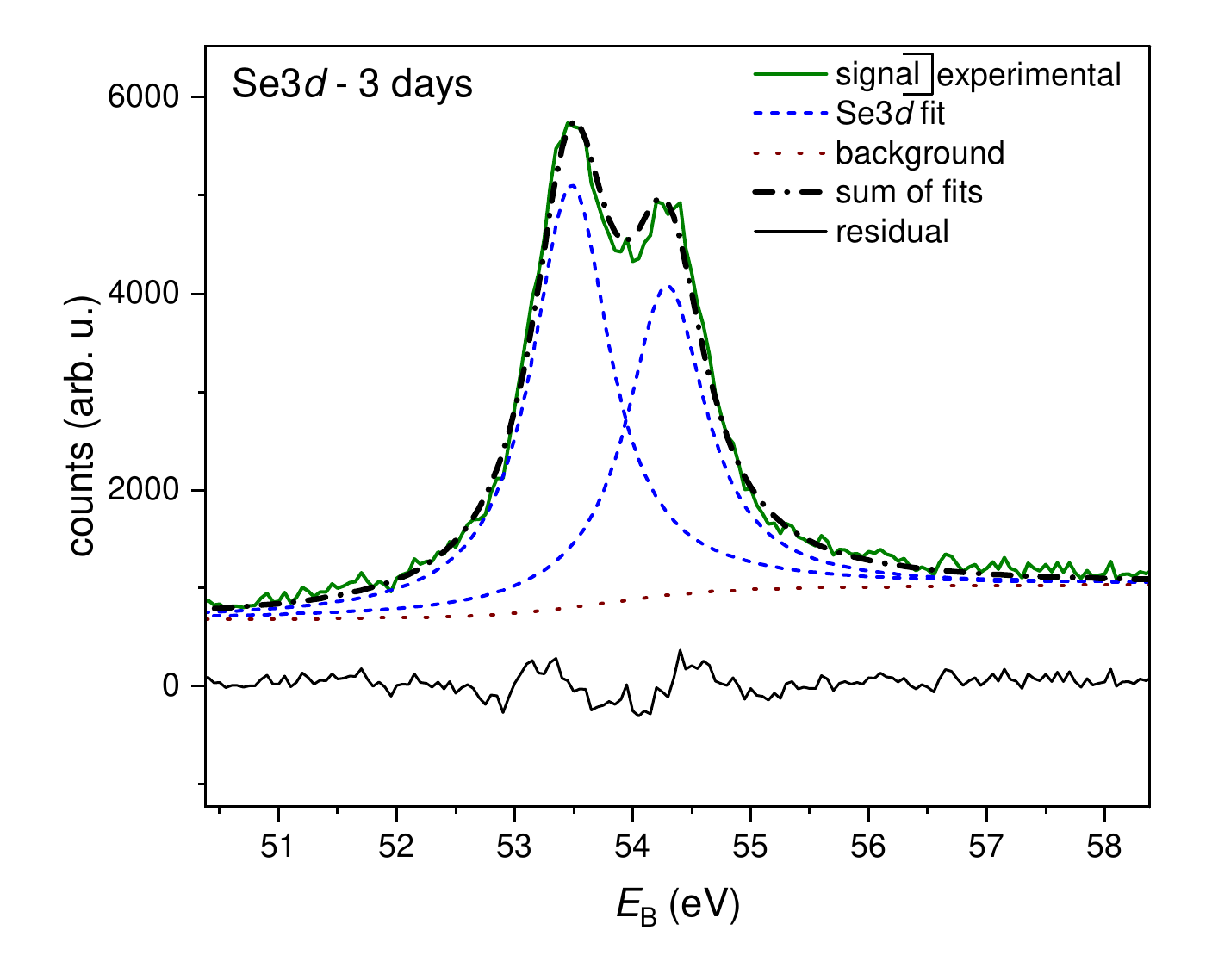}
\caption{\gls{xps} spectra, counts over binding energy, of a \pt flake after ambient air exposure for three days. Shown are the fitted Pt4$f$ spectrum (right) and Se3$d$ (left).}
\label{i_XPS_air_fit}
\end{figure}

\FloatBarrier
\subsection{Sample stability after water physisorbtion}
Upon removing the \gls{sh} from the low-$T$ He-atmosphere, there is inevitably condensation forming under ambient conditions, even if the \gls{sh} was pre-heated. When measuring the Kondo effect again in $(90,\psi)$ orientations, a significantly diminished magnitude of the Kondo resistance ($\sim50\,\text{\%}$) is observed for the resistance $(R-R(H=0))$ over applied magnetic field in \fig \ref{i_MR90psi_dim}. Potentially, physisorbed water reduces the uncompensated spin contribution at the surface. The water-condensation also resulted in an additional linear component, which is omitted in \fig \ref{i_MR90psi_dim} in order to demonstrate the diminished Kondo resistance. This points towards the sample geometry being altered, such that there are (more) transverse transport components intermixed with the longitudinal transport.

\begin{figure}[htb]
\includegraphics[width=0.6\linewidth]{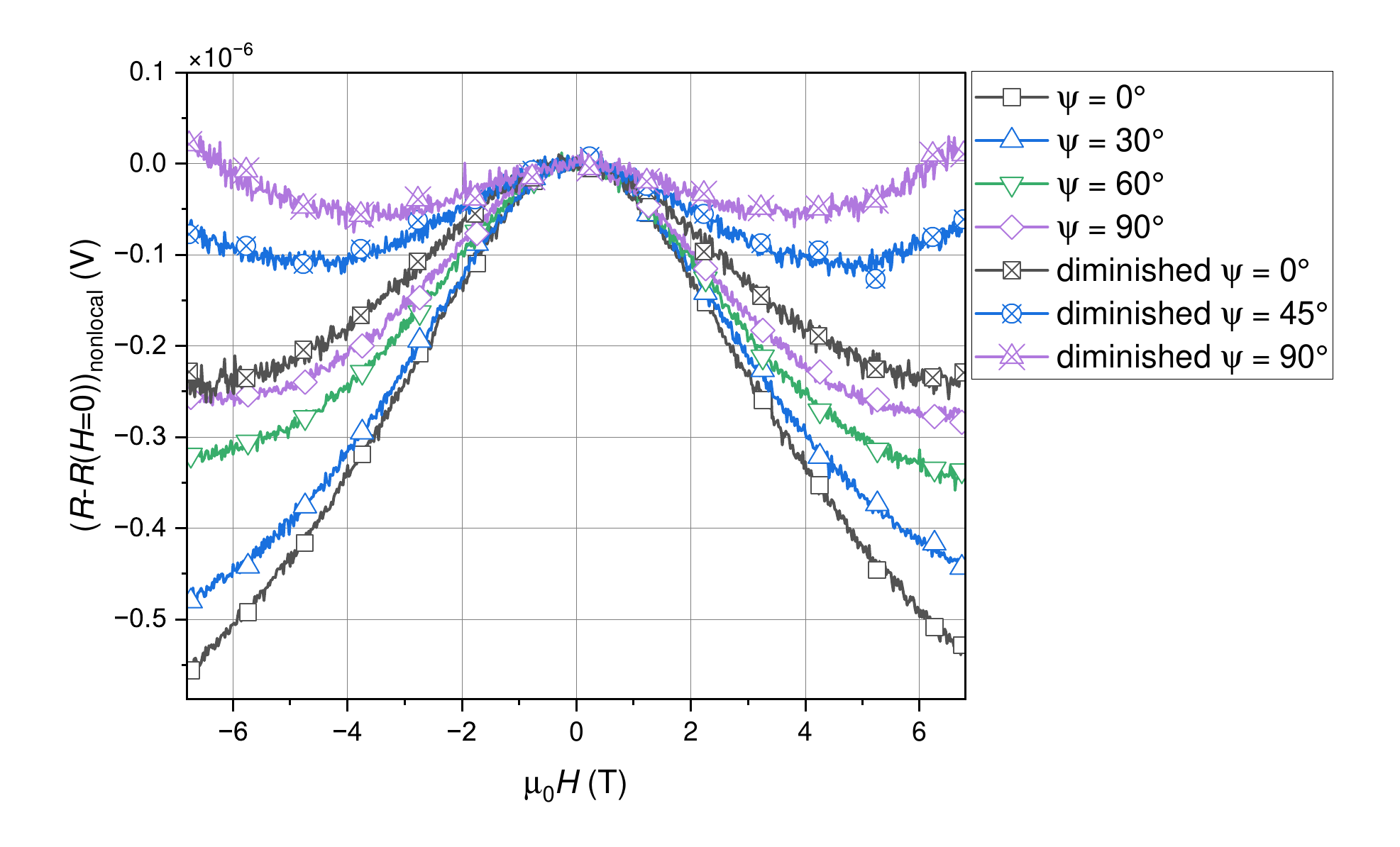}
\caption{Resistance $(R-R(H=0))$ over applied magnetic field in $(90,\psi)$ orientations for sample B: The magnitude of the Kondo resistance is diminished when comparing the curves acquired after the sample was exposed to water-condensation (symbols with crosses) to those of the pristine sample (open symbols).}
\label{i_MR90psi_dim}
\end{figure}

\FloatBarrier
\subsection{Weak anti-localization}
The presence of \gls{wal} was previously reported for \pt \cite{LiPtSe2NLMR}. \fig \ref{i_WAL} shows a low-$\mu_0$H sweep of resistance $(R-R(H=0))$ for sample A in the $(-90,0)$ orientation at $2\,\text{K}$. A quadratic fit can satisfactorily be employed in this region, as discussed in the main text. There is no apparent signature of \gls{wal}.
\begin{figure}[htb]
\includegraphics[width=0.5\linewidth]{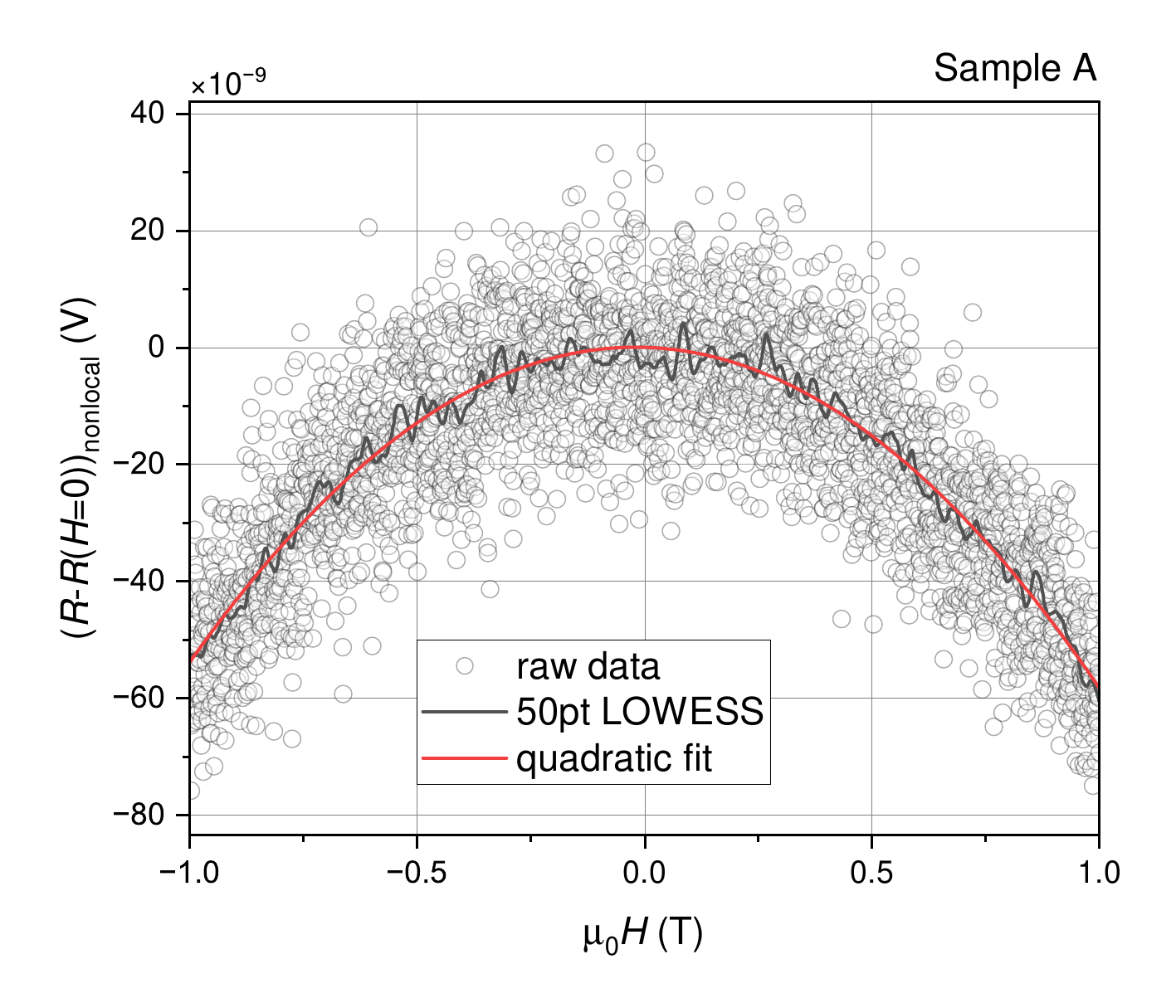}
\caption{Low-$\mu_0$H sweep of the resistance $(R-R(H=0))$ for sample A in the $(-90,0)$ orientation at $2\,\text{K}$. Besides the raw data used for fitting, \gls{lowess} with a window-size of 50 points is shown in order to high-light potential differences to the quadratic fit.}
\label{i_WAL}
\end{figure}

\FloatBarrier
\subsection{Back-gating}
The samples A,B and C offer no dedicated (top-)gating terminal. Instead, back-gating is performed by contacting the Si below the $90\,\text{nm}$ $\text{SiO}_2$. In \fig \ref{i_gate_sweep}, the effect of a gate voltage $V_G$ sweep on the resistance is shown for sample A in the $(90,0)$ orientation at $2\,\text{K}$ at different magnetic fields. Either the gating distance of $90\,\text{nm}$ in conjunction with the metallic nature of \pt greatly restrict the gating efficacy, or the density of states around $E_\text{F}$ is approximately constant. When sweeping the magnetic field at constant $V_G$, the applied $V_G$ manifests as an offset to the resistance, depicted in \fig \ref{i_MR_gate}. The gating does not qualitatively alter the \gls{mr} shape, which would be expected, if $V_G$ had managed to significantly shift the Fermi level closer or further from the energy of the Pt-vacancy spin states.

\begin{figure}[htb]
\includegraphics[width=0.4\linewidth]{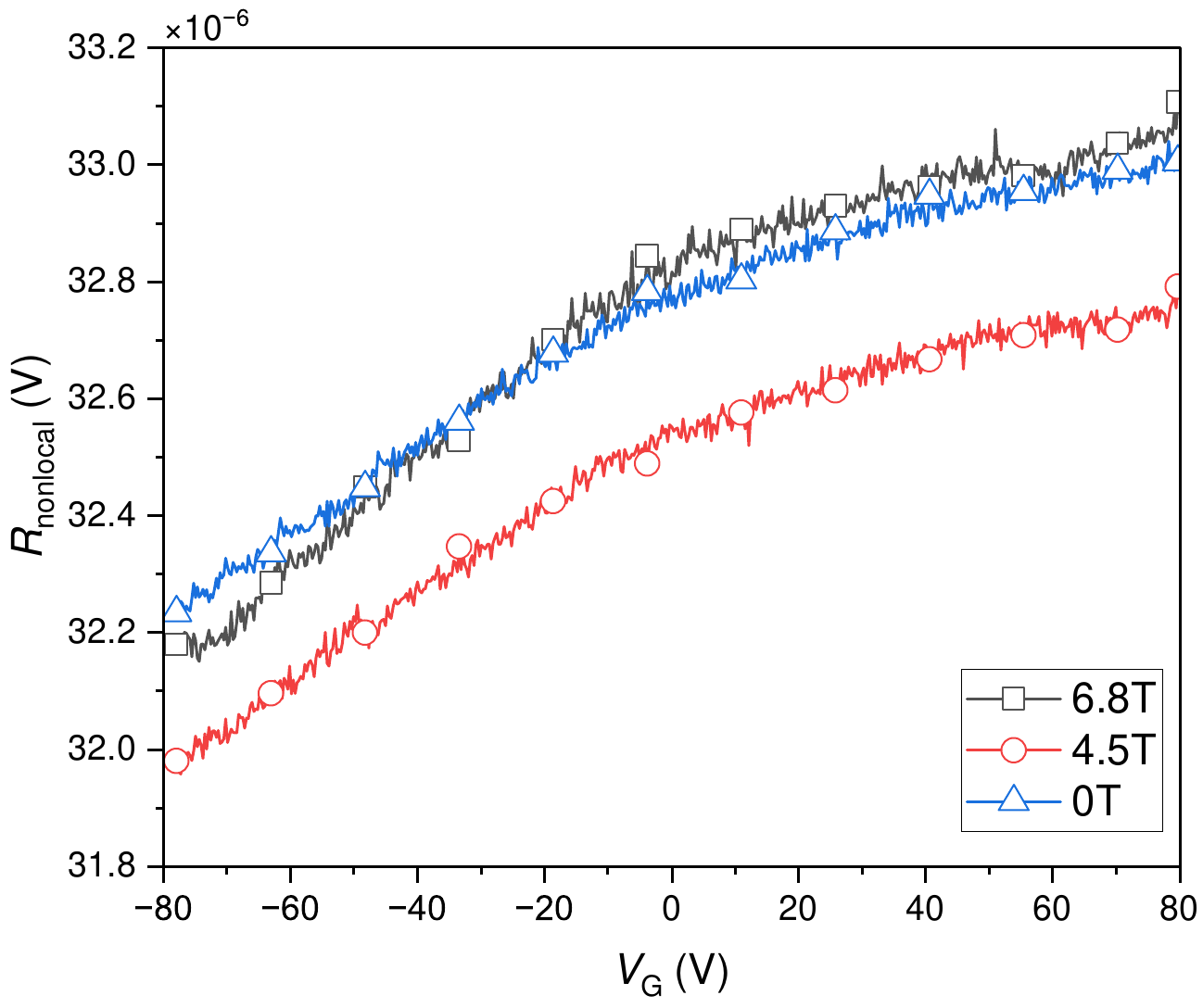}
\caption{Resistance over back-gated gate voltage for sample A in the $(90,0)$ orientation at $2\,\text{K}$. Gate voltages of up to $\pm 80\,\text{V} = V_G$ only slightly affect the resistance.}
\label{i_gate_sweep}
\end{figure}
\begin{figure}[htb]
\includegraphics[width=0.5\linewidth]{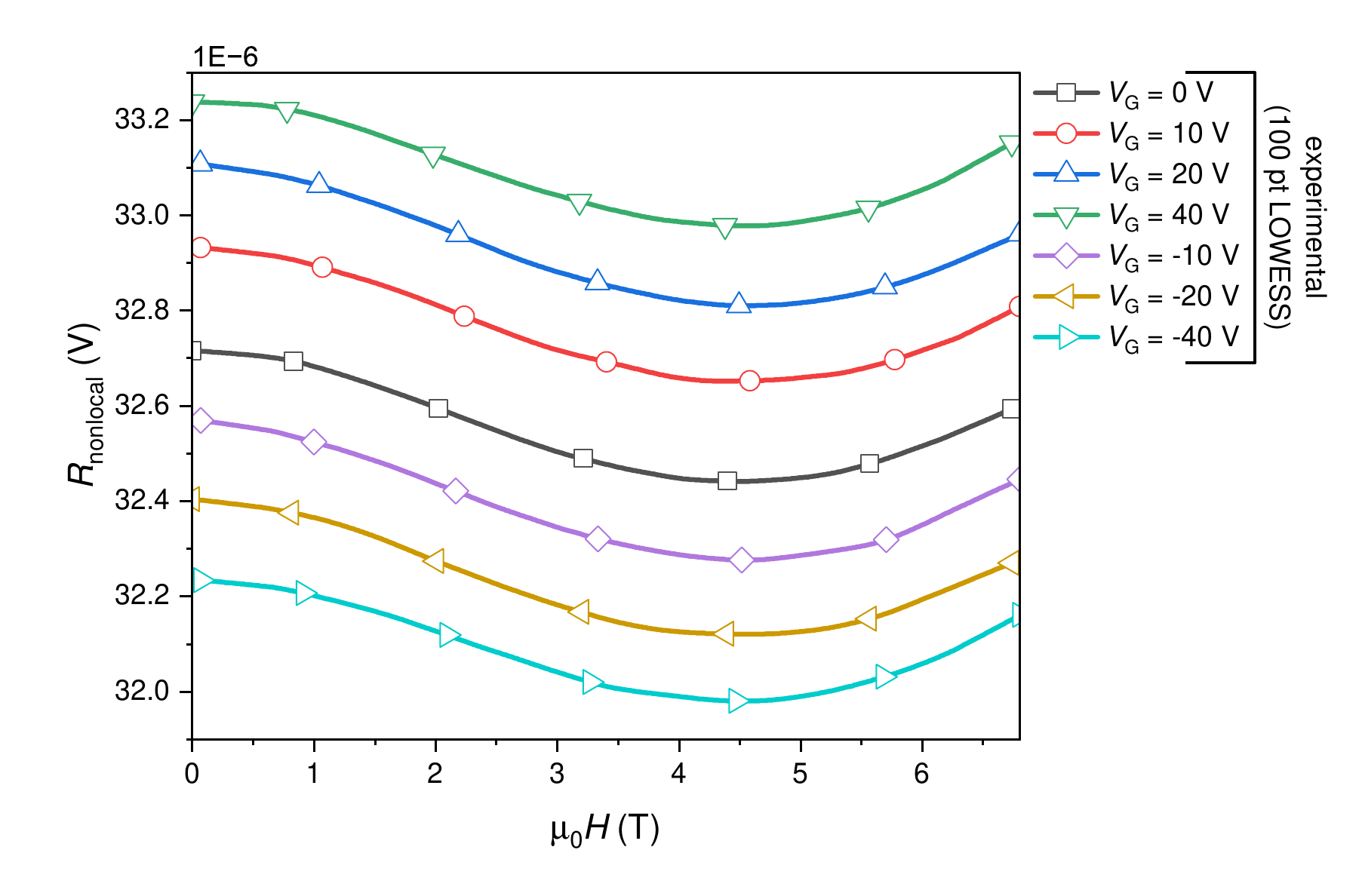}
\caption{Resistance over applied magnetic field at specific back-gating voltages for sample A in the $(90,0)$ orientation at $2\,\text{K}$. Gate voltages of up to $\pm 40\,\text{V} = V_G$ result in a (slightly saturating) offset to the \gls{mr}. $100\,\text{point}$-window \gls{lowess}.}
\label{i_MR_gate}
\end{figure}

\FloatBarrier
\subsection{Computational details}
First-principles calculations are performed within the density functional theory framework \cite{DFT} implemented in the Quantum Espresso software package \cite{QE} based on plane waves and pseudopotentials. Perdew-Burke-Ernzerhof functional and PAW pseudopotentials \cite{DALCORSO2014337} are used in all calculations. After convergence tests, the energy cutoff for the wavefunction and the charge density are set to 55 Ry and 500 Ry. The Monkhorst-Pack of 16 × 16 × 8 mesh for k-point sampling is used in calculations. The bandstructure is calculated on 416 k-points along $\mathit{\Gamma}$-$M$-$K$-$\mathit{\Gamma}$ direction. In order to simulate \gls{2d} structure, a vacuum of 20 Å is added along the z-direction to avoid interactions between the layers. Geometry optimization of atom positions is performed using the Broyden–Fletcher–Goldfarb–Shanno algorithm, with criteria for maximum allowed forces between atoms of $10^{-6}\,\text{Ry/Å}$ as the generalized gradient approximation functionals do not take into consideration long range forces as the \gls{vdw} force, Grimme-D2 \cite{grimme2006semiempirical} correction is included to obtain more accurate lattice constants and forces. A multilayer \pt model is adopted, which consist of a four-layer thick 3 × 3 supercell. The Pt defects are placed in the topmost layer and a vacuum region of 25 Å is introduced to separate the periodic images. Band unfolding from the superstructure is performed using an unfold-x program \cite{PacileUnfoldX}.

\FloatBarrier
\subsection{Supplemental figures}
The resistance $(R-R(H=0))$ over applied magnetic field in the $(0,0)$ orientation is given in \fig \ref{i_MR_0_0_BC} a) and b) for samples B and C respectively.
\begin{figure}[htb]
\includegraphics[width=0.4\linewidth]{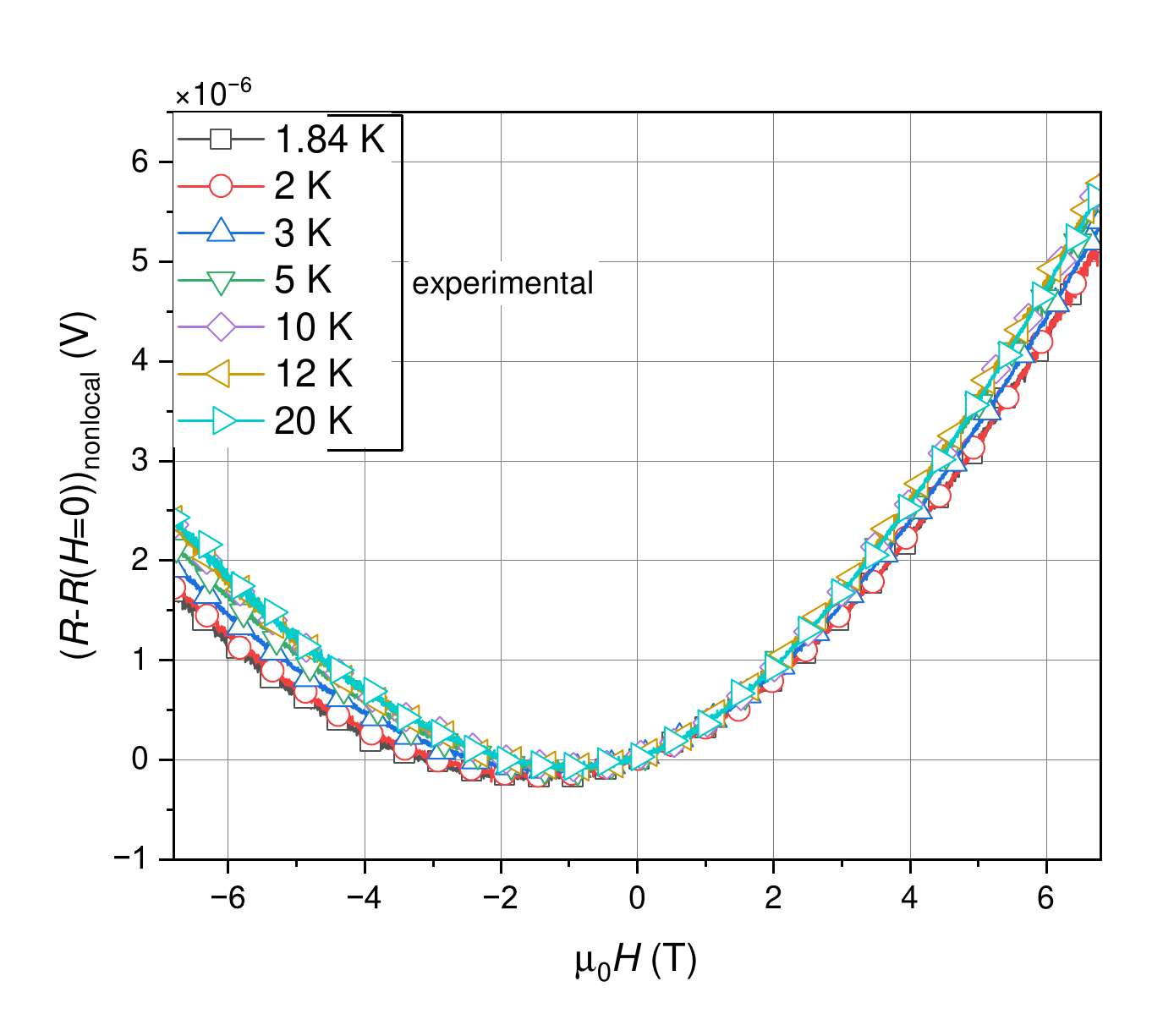}
\stackinset{c}{-0.4\linewidth}{c}{0.3\linewidth}{a)}{\hspace{0pt}}
\includegraphics[width=0.4\linewidth]{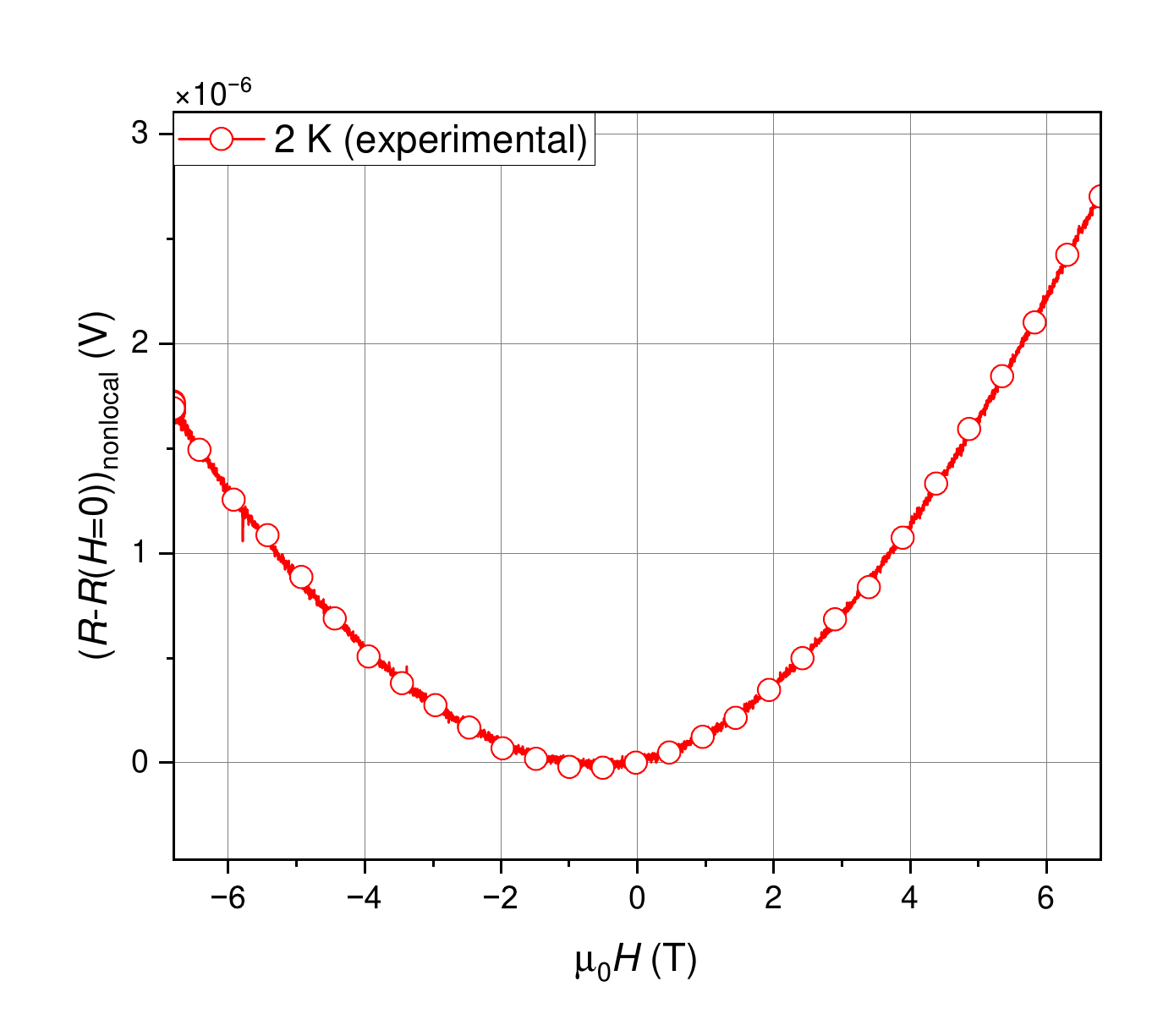}
\stackinset{c}{-0.4\linewidth}{c}{0.3\linewidth}{b)}{\hspace{0pt}}
\caption{Resistance $(R-R(H=0))$ of samples B and C over applied magnetic field in the $(0,0)$ orientation at specific temperatures. a) Sample B. b) Sample C.}
\label{i_MR_0_0_BC}
\end{figure}

The resistance over applied magnetic field (without resampling or correcting for linear contributions) for sample B in the $(90,\psi)$ orientations is depicted in \fig \ref{i_MR90_psi_raw}.
\begin{figure}[htb]
\includegraphics[width=0.5\linewidth]{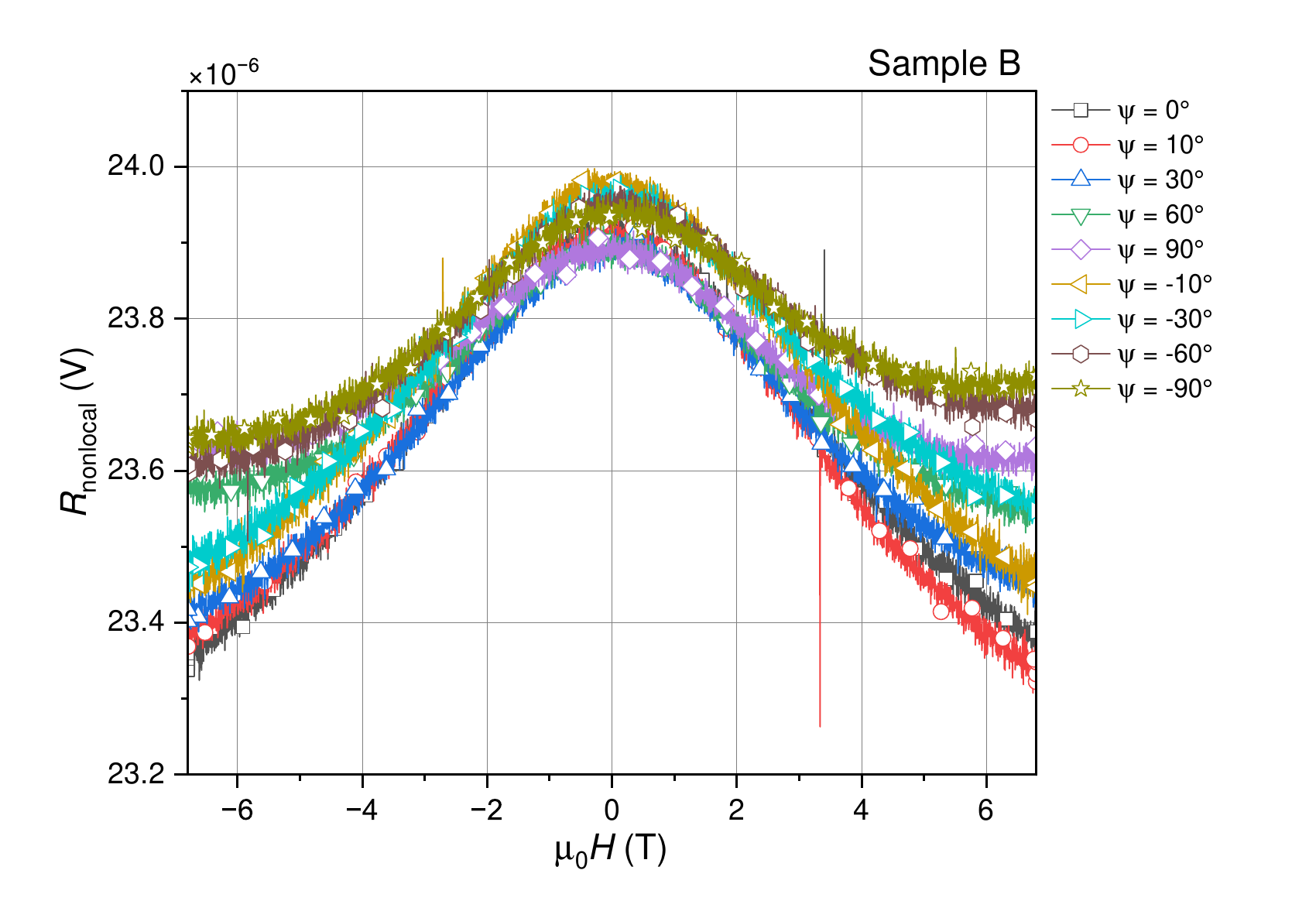}
\caption{Raw data of resistance over applied magnetic field for sample B at $2\,\text{K}$ in the $(90,\psi)$ orientations, with $\psi$ in the range of $-90\,\degree \leq \psi \leq 90\,\degree$. The offset occurring between orientations is attributed to the shifting of the cables which connect the sample contacts to the lock-in amplifier and/or the strain of the cables being altered during rotation.}
\label{i_MR90_psi_raw}
\end{figure}

\FloatBarrier
\newpage
\subsection{\label{add_tab}Supplemental tables}
Fitting parameters of the $\text{ZFC}$ data are given in \tab \ref{t_ZFC_param}. The \gls{zfc} curves being linear for most of the temperature range and the cooling rate being lower at higher temperatures would give an unwarrantedly high least-squares-method importance to this regime and lead to inferior fitting of the low-$T$ region. The fitting is thus performed on the data resampled every $0.1\,\text{T}$ to achieve uniform weighting regardless of cooling rate. The fitted temperature range is $T\epsilon[0,50]\,\text{K}$. The linear regime temperature is $T_\text{lin} \approx 80\,\text{K}$ and therefore not evaluated.

\begin{table}[htb]
\begin{tabular}{cc|ccccc}
sample & thickness (nm) & $R_\text{res}$ ($\mu$V) & $R_\text{K0}$ ($\mu$V) & $T_\text{K}$ (K)     & $S$\\ \hline
A      & 18             & $30.0\pm 0.1$           & $4.2\pm 0.2$      & $9.7\pm 0.3$ & $0.35\pm 0.03$\\ 
B      & 20             & $22.44\pm 0.05$         & $1.79\pm 0.08$    & $8.8\pm 0.2$  & $0.22\pm 0.02$\\ 
C      & 26             & $15.24\pm 0.08$           & $1.2\pm 0.1$      & $12\pm 1$     & $0.22\pm 0.03$\\ 
\end{tabular}
\caption{Fitting parameters of the \glspl{zfc} for the considered samples data using \eq \ref{eq_ZFC}. Prefactors $A_F$ and $A_\text{ph}$ omitted.}
\label{t_ZFC_param}
\end{table}

\subsection{Supplemental equations}

In order to fit the \gls{rt} at temperatures of $T>T_\text{lin}$, the regime of $T<T_\text{lin}$ is extended linearly and $C_1$-continuously. \eq \ref{eq_RSM_full} gives the full expression for the semimetallic resistance $R_\text{SM}$ at arbitrary temperatures. 

\begin{equation}
    R_\text{SM}(T) = R_\text{res}
    +A_\text{F} {\begin{cases}
        T^2,&T\leq T_\text{lin}\\ 
        2\,T_\text{lin} T - T_\text{lin}^2,& T > T_\text{lin}
    \end{cases}}
    +A_\text{ph} {\begin{cases} 
        T^5,&T\leq T_\text{lin}\\
        5\,T_\text{lin}^4 T - 4\,T_\text{lin}^5,& T > T_\text{lin}
    \end{cases}}
    \label{eq_RSM_full}
\end{equation}

The expanded form of the general description of the longitudinal resistance (\eq \ref{eq_master_simple}) is:

\begin{equation}
\begin{split}
    R(H,T,\theta,\psi)&= R_\text{res}
    +A_\text{F} {\begin{cases}
        T^2,&T\leq T_\text{lin}\\ 
        2\,T_\text{lin} T - T_\text{lin}^2,& T > T_\text{lin}
    \end{cases}}
    +A_\text{ph} {\begin{cases} 
        T^5,&T\leq T_\text{lin}\\
        5\,T_\text{lin}^4 T - 4\,T_\text{lin}^5,& T > T_\text{lin}
    \end{cases}} \\
    &+ R_{\text{K0}}\frac{1}{2}\left(1-\frac{\ln({T/T_\text{K}})}{\sqrt{\ln({T/T_\text{K}})^2+S(S+1)\pi^2}}\right)
    \left( 1-\mathfrak{B}_J\left(\frac{g \mu_B \mu_0 H}{k_B (T+T_\text{K})}\right)^2\right) \\
    &+\left( M_{\bot}\left|\mu_0 H\right|^{p_\bot} +k_\bot \mu_0 H\right)(\cos\theta)^2 (\cos\psi)^2
    +\left( M_{\parallel}\left|\mu_0 H\right|^{p_\parallel} +k_\parallel \mu_0 H\right)(\sin\psi)^2.
    \label{eq_master}
\end{split}
\end{equation}

\clearpage
\end{widetext}
\end{document}